\documentclass[twocolumn, notitlepage]{revtex4-2}
\usepackage{blindtext}
\usepackage{graphicx}
\usepackage{dcolumn}
\usepackage{bm}
\usepackage{amssymb}
\usepackage{appendix}
\usepackage{url}
\usepackage{color}
\usepackage{xcolor}
\usepackage[T1]{fontenc}
\usepackage{mathrsfs,amsfonts,dsfont}
\usepackage{nccmath}
\usepackage{amstext}
\usepackage{mathtools}
\usepackage{enumitem}
\graphicspath{{graphics/}}
\usepackage{rotating} 
\usepackage[english]{babel}  
 
\usepackage{wrapfig}
\usepackage{lipsum} 

\usepackage{braket}
\usepackage{bbm} 
\usepackage[colorlinks=true,citecolor=blue,linkcolor=blue]
{hyperref}

\usepackage{amsthm}

\begin{document}

\title{Multipartite multiplexing strategies for quantum routers}
\author{Julia A. Kunzelmann, Hermann Kampermann, Dagmar Bruß}
\affiliation{Institute for Theoretical Physics III, Heinrich Heine University Düsseldorf, D-40225 Düsseldorf, Germany }
\begin{abstract}
This work explores the important role of quantum routers in communication networks and investigates the increase in efficiency using memories and multiplexing strategies. Motivated by the bipartite setup introduced by Abruzzo et al. (2013) for finite-range multiplexing in quantum repeaters, we extend the study to an \emph{N}-partite network with a router as a central station. We present a general protocol for \emph{N} parties after defining the underlying matching problem and we calculate the router rate for different \emph{N}. We analyze the improvement due to multiplexing, and analyze the secret key rate with explicit results for the tripartite network. Investigating strategic qubit selection for the GHZ measurements, we show that using cutoffs to remove qubits after a certain number of rounds and consistently combining qubits with the lowest number of storage rounds leads to an optimal secret key rate. 
\end{abstract}
\maketitle

\section{Introduction}

Quantum communication is a major field of research in quantum information theory. An essential area is here the generation of secret keys, which can be used in cryptography. In quantum key distribution (QKD), such keys are generated between two parties \cite{BB84_Original, Ekert}. The generalization to $N$ parties is called conference key agreement (CKA) \cite{CKA, NBB84}. Usually, photons are used to distribute a key which limits the distance to about 150 km \cite{Repeater} as photon losses scale exponentially with distance. Quantum repeaters are needed to overcome this problem \cite{Repeater1, QuantumRepeater2}. In an intermediate repeater station, the entangled state between the remote parties is established by performing a Bell state measurement (BSM). So far, quantum repeaters connecting two parties have been investigated, either without memories \cite{Exprepeater} or with memories \cite{MDIQKD}. In the latter case, additional multiplexing can be used to perform parallel independent Bell state measurements, thus increasing the generation rate of entangled states between the parties \cite{Multi_Original, Multiplexing} per round. 
These states can e.g. be used for quantum key distribution such as the BB84 protocol \cite{BB84_Original} or (measurement) device-independent QKD protocols \cite{MDIQKDOriginal, Ekert}.

In this work, we introduce a generalization to a quantum router that connects $N$ parties in a star graph, where the quantum router is the central node. This central station is used to distribute multipartite entanglement between the parties in larger networks \cite{Epping}. 
Similar network structures have been analyzed in quantum switches with or without buffer, where the goal is to connect $N \leq k$ of the \emph{k} users via an entangling measurement. This setup has been investigated numerically \cite{Coopmans_2021} and also analytically using Markov chains \cite{Vardoyan_2021}. 
Here, we deal with the distribution of entangled GHZ states between $N$ parties by performing GHZ measurements within the router. We additionally include quantum memories for multiplexing in the quantum router by generalizing the protocol from \cite{Multiplexing}. 
We define the underlying $N$-dimensional matching problem, discuss suitable algorithms for different network sizes, and analyze the router rate. In a second step, we analyze such networks in the context of conference key agreement. Our setup can be seen as a generalization of the MDI-QKD protocol with quantum memories from \cite{MDIQKD} to more than two parties.
Compared to previous work about CKA in star graphs \cite{Multiplexing1}, we additionally analyze the effect of quantum memories and the use of multiplexing. We calculate the quantum bit error rates for the BB84 protocol with \emph{N} parties and use this to determine the asymptotic secret fraction and the secret key rate. The focus is on examining different matching strategies to select the qubits for the GHZ measurements to maximize the key rate. 
\\ \\
\indent
The paper is structured as follows. In Sec. \ref{sec:QR} we present the $N$-partite network with the quantum router as the central element and explain the entanglement distribution among all parties. 
In Sec. \ref{sec:N-party-mult} we introduce multipartite multiplexing and define the related matching problem from graph theory. We further focus on the router rate, i.e. the rate with which entangled states can be created in each round of the protocol. Router rates for different setups are calculated. 
Finally, we consider conference key agreement and determine secret key rates for tripartite networks in Sec. \ref{sec:CKA}. 
We further analyze different strategies for minimizing the influence of memory decoherence for the tripartite network. We end in Sec. \ref{sec:Conclusion} with our conclusion and outlook.

\section{Quantum router with memories in \emph{N}-partite star graphs}
\label{sec:QR}

\begin{figure}
    \includegraphics[scale=0.27]{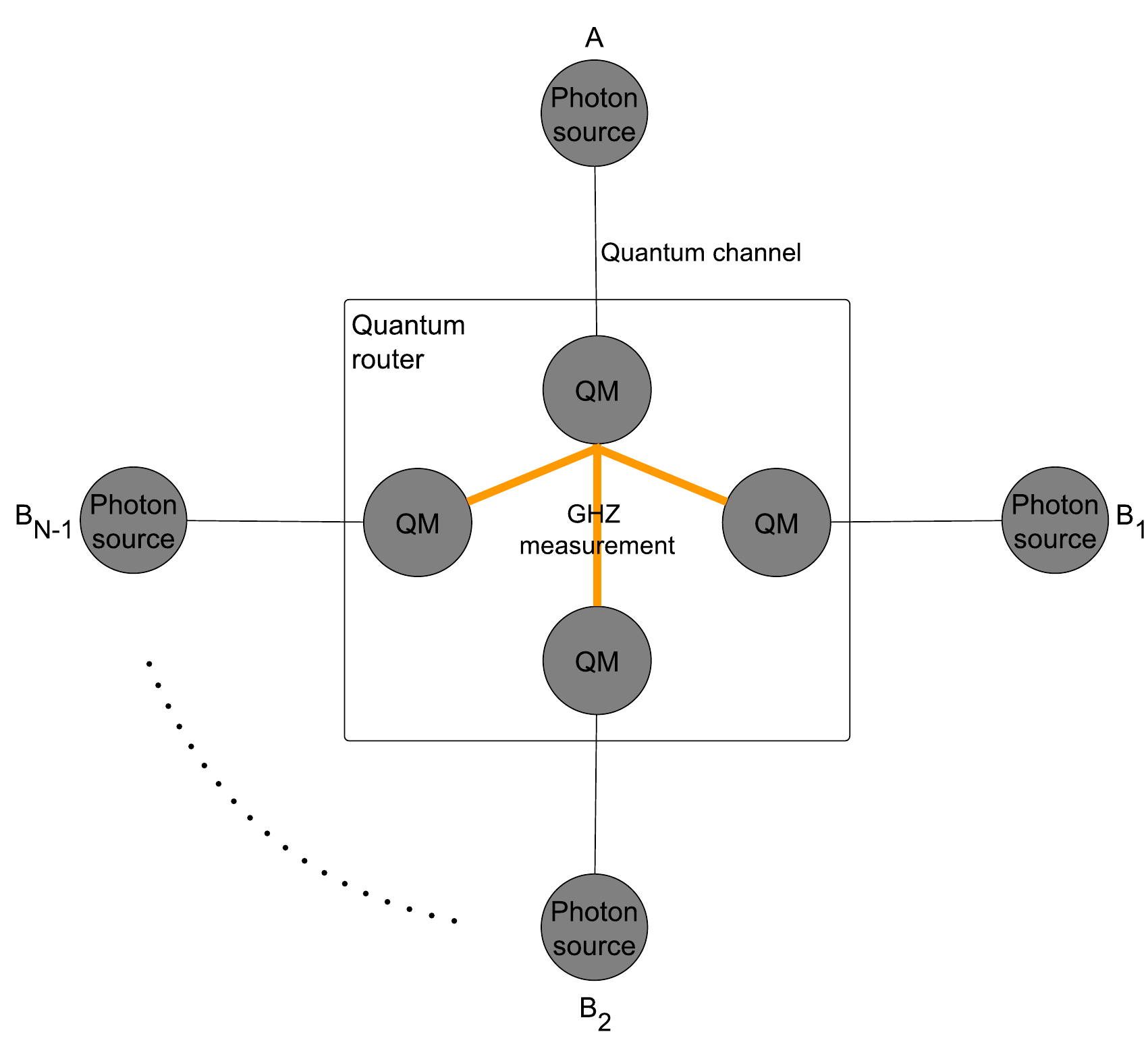}
    \caption{General setup of a quantum router with quantum memories (QM) that is connected to $N$ parties. All parties are placed at equal distances around the router. The photon sources of the parties each produce a Bell state, of which one qubit is held locally while the second qubit is sent to the central station. A GHZ measurement between all parties is performed.  }
    \label{fig:MDIQKD_with_memories}
\end{figure}

\begin{figure}
    \includegraphics[scale=0.9]{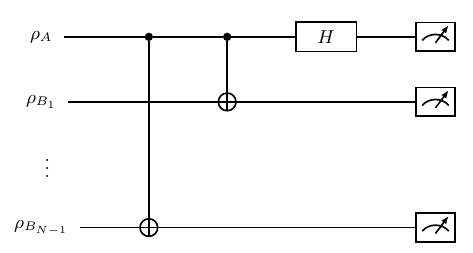}
 \caption{Quantum circuit for performing the entangling GHZ measurement. Gates are only applied to the qubits stored in the memories. } 
\label{fig:GHZ}
\end{figure}

We first generalize the concept of a quantum repeater to a quantum router in a network of \emph{N} communicating parties that are located at equal distances around the router. The scenario with unequal distances of the parties could be analyzed in an analogous way.
The general setup of such a star-shaped network is shown in Fig. \ref{fig:MDIQKD_with_memories}. 
The entanglement distribution is performed in the following steps: 
\begin{enumerate}
    \item Each party prepares a Bell state $|\phi^+ \rangle = \frac{1}{\sqrt{2}} \left( |00\rangle + |11\rangle \right)$ and sends one qubit via the quantum channel to the quantum router. The second qubit is held locally by each party. Qubits that successfully arrive at the router are stored in a memory. 
    \item In each round, in which some memories of all parties are filled, a GHZ measurement is performed (Fig. \ref{fig:GHZ}), where one party (here party A) has a special role, providing the control qubit for the CNOT gates and performing the Hadamard gate. The measurement outcome is announced to all parties. 
    Memories, whose stored qubits are included in a GHZ measurement are reset for the next round. Filled memories, which are not included in a GHZ measurement remain filled for the next round.
    \item Depending on the measurement outcome, the parties perform a phase flip (party A) or bit flip (parties B$_i$), if necessary, in order to obtain the desired GHZ state $|GHZ \rangle_N = \frac{1}{\sqrt{2}} \left( |0\rangle^{\otimes N} + |1\rangle^{\otimes N} \right)$. 
\end{enumerate}
We call these three steps together "one round". In the following, the label of the round is denoted by $s$. Any number of rounds can be performed, each containing one attempt of entanglement generation. 
By including $m>1$ memories per party, the probability that each party has at least one filled quantum memory in a round can be increased. This means that the number of distributed states in a single round increases to a maximum of $m$. This so-called multiplexing is examined in detail in Sec. \ref{sec:N-party-mult}.
\par
As a figure of merit, we define the router rate as the number of successful GHZ measurements per memory per round averaged over the whole running time up to a current round $s_c$, namely:
\begin{align}
\label{eq:Rc}
    R(s_c) = \frac{1}{s_c} \sum_{s = 1}^{s_c} \frac{\langle l \rangle (s)}{m}
\end{align}
Here, $\langle l \rangle (s)$ is the average number of GHZ measurements in round \emph{s}. 
During the storage process, the qubits are subject to noise, which is analyzed in the following.

\subsection*{Noise model for the memories} 
\label{sec:noisy}
We model noise affecting the stored qubits by depolarization. Starting the storage in round $s_0$ with a given quantum state $\rho_0$, the depolarized state in round $s$ has the form:
\begin{align}
    \label{eq:whitenoise}
    \rho\left( s-s_0 \right) = p(s-s_0) \rho_0 + \frac{1-p\left(s-s_0 \right)}{2}\mathds{1}
\end{align}
The decoherence parameter $\tau$ of the memory is related to the probability of white noise (see Eq. (\ref{eq:whitenoise})) by 
\begin{align}
\label{eq:deco}
p(\delta) = e^{-\delta/\tau} 
\end{align}
with $\delta = s-s_0$  being the number of storage rounds of a qubit. Note, that the decoherence parameter $\tau$ and the number of storage rounds $\delta$ are each given by an integer. 

The bipartite states after decohering in the memories are given by:
\begin{multline}
    \rho_i^{dep} = F_i |\phi^+\rangle \langle \phi^+|  \\
    + \frac{1-F_i}{3} \left( |\phi^-\rangle \langle \phi^-| + |\psi^+\rangle \langle \psi^+| + |\psi^-\rangle \langle \psi^-| \right)
\end{multline}
with $F_i = \frac{1}{4} + \frac{3}{4} \cdot p(\delta)$ defining the fidelity of the states for the parties $i \in \{A, B_1, B_2, ...\}$. 
The total input state $\rho_{AB_1\hdots B_{N-1}}^{dep}$ is given by the tensor product of the noisy states provided by each party. Performing the GHZ measurement on those qubits which are stored in the memories, the parties end up sharing a GHZ diagonal state $\Tilde{\rho}^{dep}_{AB_1\hdots B_{N-1}}$ of the remaining qubits held locally. It is related to the input fidelities $F_i$ of the initial depolarized states by its GHZ diagonal elements. The fidelity of the output state is given by $\Tilde{F}_{\Tilde{\rho}^{dep}} = \langle GHZ | \Tilde{\rho}^{dep}_{AB_1\hdots B_{N-1}}| GHZ \rangle_N$. An explicit calculation for the tripartite case is given in Appendix \ref{sec:Tripartite}.

\section{\emph{N}-partite multiplexing} 
\label{sec:N-party-mult}

\begin{figure}
    \centering
    \includegraphics[scale=0.26]{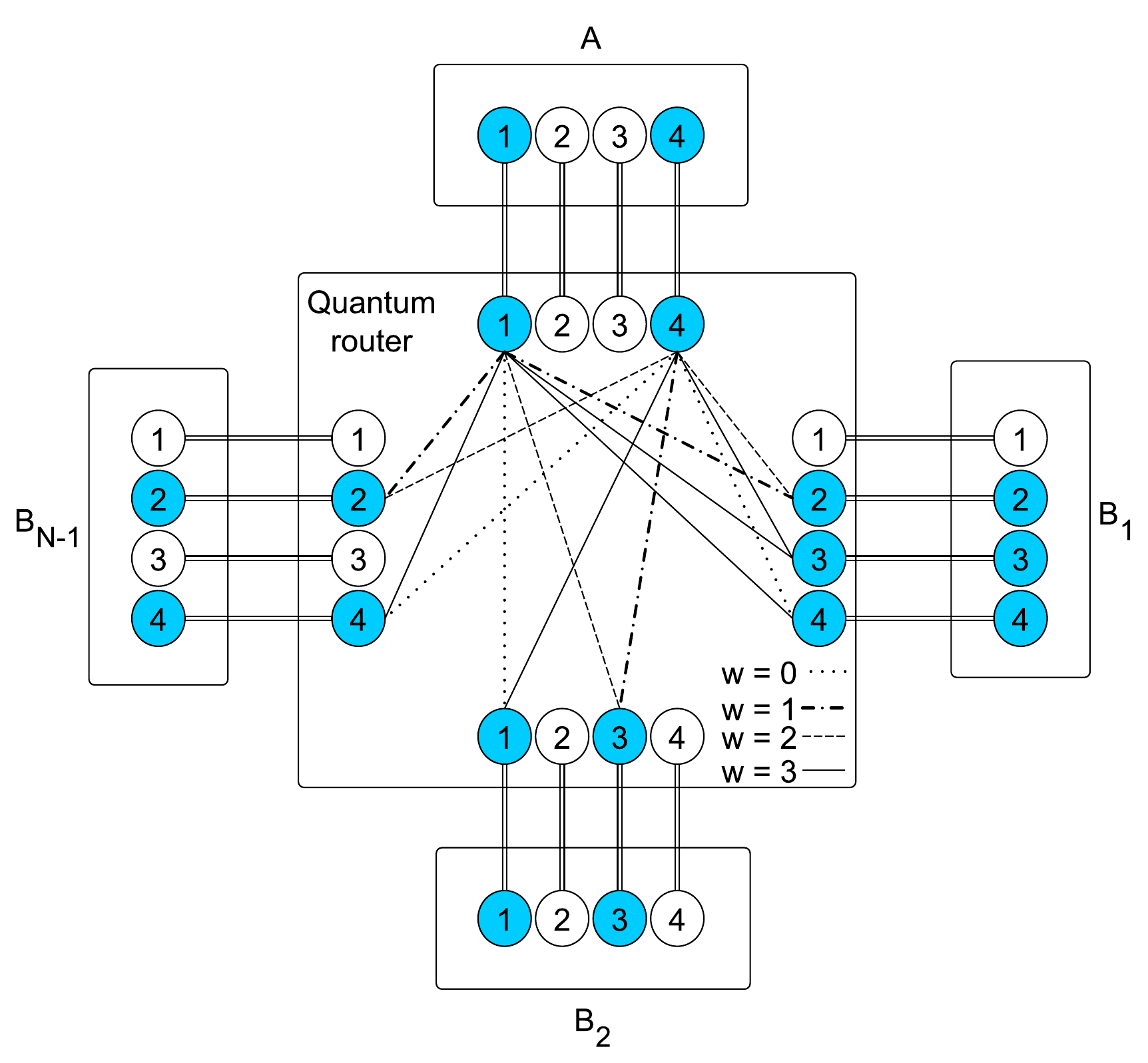}
    \caption{Generalized setup of a quantum router with multiplexing for an \emph{N}-partite star graph. Shown is an exemplary filling of the memories. Empty memories are white, while filled memories are shown in blue. The connection lengths $w$ (difference of labels of filled memories) for this example are also indicated. }
    \label{fig:NMulti}
\end{figure}

The generalized setup of a quantum router with multiplexing is shown in Fig. \ref{fig:NMulti}. 
All parties have a fixed number \emph{m} of photon sources and correspondingly \emph{m} quantum memories in the router. 
The memories within the quantum router are either empty (white in Fig. \ref{fig:NMulti}) or filled with a qubit (blue in Fig. \ref{fig:NMulti}). The number of filled memories depends on the loss rate of the channel and the probability of successful storage. 
The connection length $w$ is defined as the distance between the memories, i.e. the difference of the labels, see example in Fig. \ref{fig:NMulti}. 
For each memory configuration in every round, the protocol aims to perform the maximal number of GHZ measurements between all parties. 

\subsection{Time structure of a multiplexing protocol round }
\label{sec:protocol}
 
\begin{figure}
    \centering
    \includegraphics[scale=0.2]{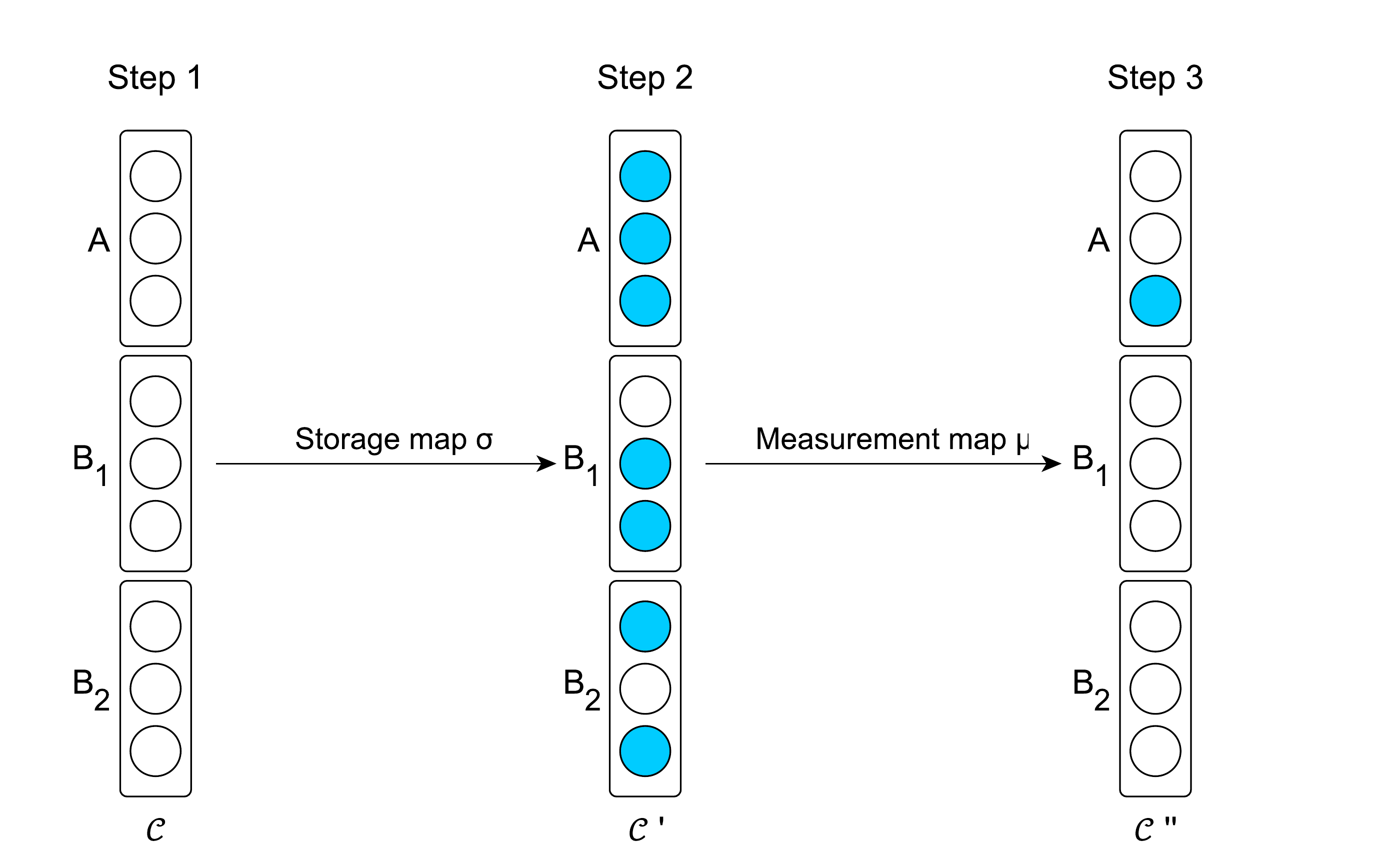}
    \caption{Temporal structure of a round in the multiplexing protocol. In each round, qubits are first sent to the router and, if they arrive successfully, are stored in the respective memories (storage map $\sigma$ leading from configuration $\mathcal{C}$ to $\mathcal{C}'$). The GHZ measurements are performed and memories that are involved in a measurement are emptied (measurement map $\mu$). The new memory configuration $\mathcal{C}''$ forms the starting configuration $\mathcal{C}$ for the following round.}
    \label{fig:MultiRound}
\end{figure}

The different configurations of the memories the quantum router goes through in one round of the entanglement distribution protocol are shown in Fig. \ref{fig:MultiRound}. 
Each memory configuration is given by a vector $\mathcal{C}$ (for details of its notation see subsection \ref{sec:RepeaterRate}), where each entry represents one memory within the router.  
Following the steps for entanglement distribution given in Sec. \ref{sec:QR}, one finds two transitions of the memory configurations: starting a round with a memory configuration $\mathcal{C}$, it changes to the intermediate configuration $\mathcal{C}'$ after sending and storing one part of the Bell pairs prepared by the parties. The qubit arrives with probability $\eta$ and is then heralded and stored by the quantum router.   
Based on the configuration $\mathcal{C}'$, a maximal number of GHZ measurements $l$ is performed.
In each round, the router reports whether the measurement was successful. If so, the measurement results as well as the information on which memories were involved in the measurement are communicated. The final memory configuration after resetting the used memories, is given by $\mathcal{C''}$. This memory configuration is kept for the next round.

In the following, the focus will be on the choice of the memories included in a GHZ measurement. Finding a combination that maximizes the number of GHZ measurements $l$ per round corresponds to the problem of matching from graph theory that is introduced in the following subsection.

\subsection{The matching problem}
\label{sec:match}

\begin{figure}
    \centering
    \includegraphics[scale=0.18]{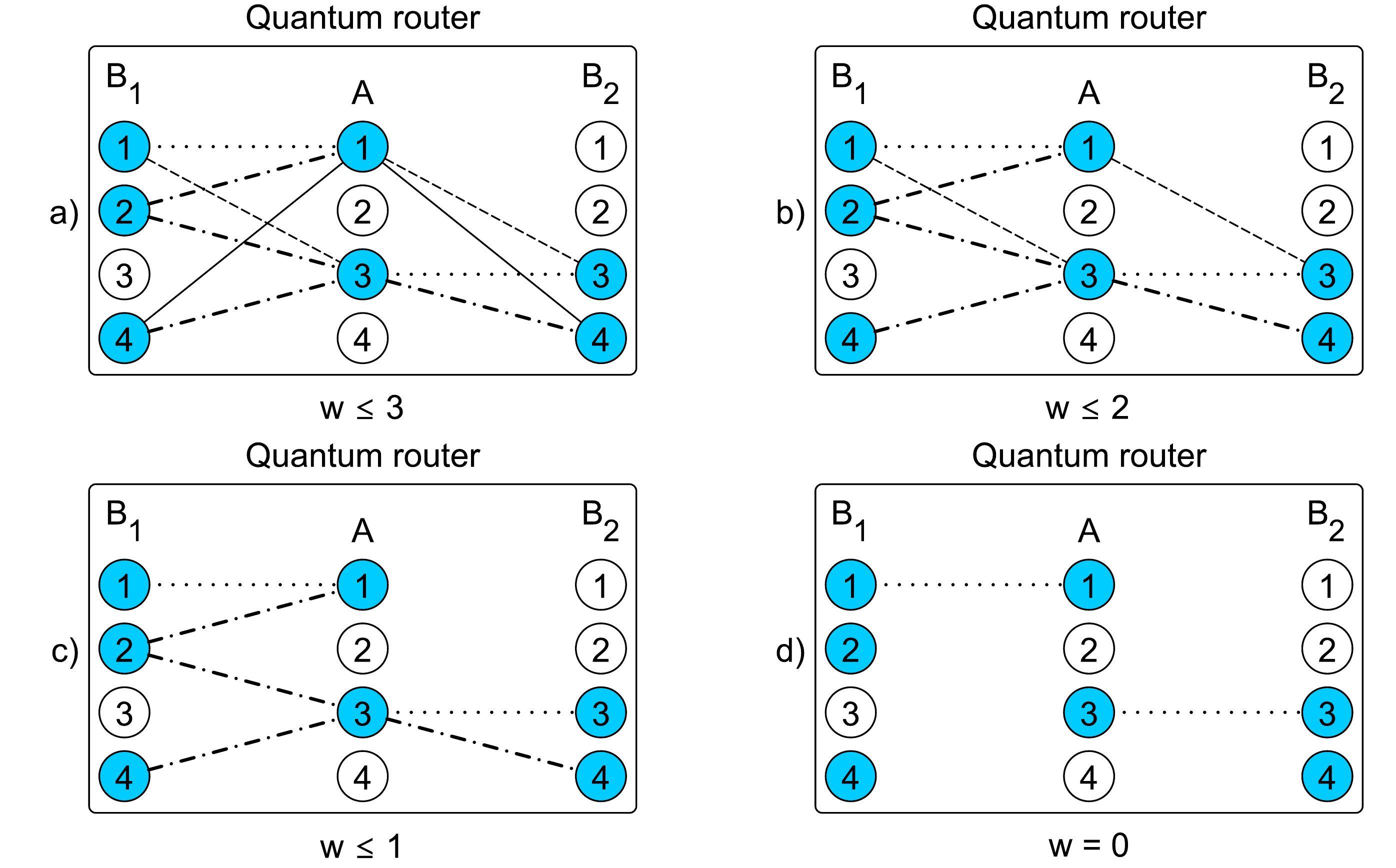}
    \caption{Example of a tripartite graph in the quantum router for different choices of the maximal connection length $w$. Part a) shows the graph for full-range multiplexing such that any node of one subset can be connected to any node of another subset. This leads to a matching with $l = 2$ contained hyperedges. In b) and c) the maximal connection length is reduced to two and one, respectively, which corresponds to finite-range multiplexing. Here, a matching with $l = 2$ and $l = 1$ contained hyperedges, respectively, arises. Finally, in d), only parallel connections are allowed ($w=0$) so that some filled memories are no longer considered in the matching. In this case, no matching is found. Solid lines represent connections with $w=m-1=3$ (maximal), dashed lines represent $w=2$, mixed (dashed-dotted) lines represent $w=1$ and dotted lines show connections belonging to $w=0$. }
    \label{fig:matching_w_dep}
\end{figure}

The memories of the quantum router form an \emph{N}-partite graph $G=(V, E)$ consisting of nodes \emph{V} (here the filled memories) and edges \emph{E} (the connectability between the memories). The set of nodes $V$ is divided into \emph{N} pairwise disjoint subsets $V_1, V_2, ..., V_N$ which result from the allocation of the memories to the different parties. An edge always connects two nodes from different subsets, i.e. $E=\{\{v_i, v_j\}|v_i \in V_i \wedge v_j \in V_j \text{ for } i\neq j \}$. 
Since the goal of the router is to perform a GHZ measurement
between party A and each B$_i$, we define hyperedges (sets of nodes) which need to fulfill the following properties:
\begin{enumerate}
    \item Each hyperedge always consists of a set of $N$ nodes from different subsets, i.e. $T = \{ \{v_1, \hdots, v_N \} | v_1 \in V_1, \hdots, v_n \in V_N \} \subseteq V_1 \times  \hdots \times V_N$, containing exactly one memory per party.  
    \item The vertices are connected in such a way that each B$_i$ has exactly one edge to party A but no edges to other B$_i$ or the own subset since this is fixed by the GHZ measurement circuit (see Fig. \ref{fig:GHZ}). In general, nodes can appear in several hyperedges. Party A can be fixed for the whole protocol or it can be chosen individually in each round. 
    \item In the case of restricting the connection length $w$, edges are only allowed to be drawn between two nodes fulfilling that constraint. 
\end{enumerate} 
In every protocol round the graph in the quantum router is constructed based on the given memory configuration.  
Memories only contribute to the graph if they are filled. Hyperedges are drawn between the memories such that they fulfill the properties defined above. 
The graph construction for a fixed memory configuration but different connection lengths $w$ is exemplary shown in Fig. \ref{fig:matching_w_dep}.   
For example, we find the following hyperedges for the graph in c): $\{ \{ 2,3,3\}, \{ 2,3,4\}, \{ 4,3,3\}, \{ 4,3,4\} \} $.
Here, we fix party A for the whole protocol, since the increased rate due to a dynamic choice of a different Alice in each round is small compared to increasing the connection length. Additionally, it is not clear, whether this can be realized easily in an experimental setup.

Since the goal of the multiplexing scheme is to perform a maximal number of GHZ measurements per round, we want to find a set of pairwise disjoint hyperedges where the set has maximal cardinality. The condition of being pairwise disjoint follows from the fact, that a stored qubit cannot be used in two different GHZ measurements.  
A set of hyperedges in which no two hyperedges share a common node is called a matching. A set of maximum cardinality concerning the number of contained hyperedges is called maximum matching.  
Note, that more than one maximum matching may exist. 
In the previous example of Fig. \ref{fig:matching_w_dep} c), each allowed hyperedge is a valid matching with a maximum cardinality of one. A larger set of hyperedges cannot be formed, since all hyperedges share Alice's third node as a common node.  

The computational problem from graph theory is formally defined as follows: 
\begin{center}
\begin{tabular}{ll}
\hline \hline 
\multicolumn{2}{c}{\textsc{Maximum N-dimensional matching} } \\
\hline 
\textbf{Given:}  & An $N$-partite graph instance and all valid \\
& hyperedges $T \subseteq V_1 \times V_2 \times \hdots \times V_N$   \\
\textbf{Question:}  & What is a valid maximum matching $M \subseteq T$? \\ 
\hline \hline
\end{tabular}
\end{center}
\bigskip
The maximum number of hyperedges (defining the number of GHZ measurements \emph{l} that can be performed in each round) is limited by the given graph instance in every round, namely by the number of existing edges between the nodes (given by the degree $\text{deg}(v_j)$ of each node $v_j$), and by the number of filled memories $n_k$ per party \emph{k}:
\begin{align}
\label{eq:maxk}
    \min_{k \in \{A, B_i\} } \{ \left| \{v_j | \text{deg}(v_j) >0 \wedge v_j \in V_k \} \right| \} \leq l \leq \min_{k \in \{ A, B_i \} } \{n_k\} .
\end{align}
Remember that the degree of a node may depend on the maximal connection length $w$.
For full-range multiplexing it holds $l = \min_{k \in \{ A, B_i \} } \{n_k\} $. 
Finding a matching in this case as well as for $w=0$ is trivial. It turns out, however, that the general problem is contained in the class $\mathcal{NP}$ when restricting the connection length \emph{w} (compare to the hardness results of the general $N$-dimensional matching problem considered in \cite{Karp, ProofNP}).  
Nevertheless, it is possible to construct algorithms that solve the previously introduced maximum \emph{N}-dimensional matching problem for small graphs if the total number of memories does not become too large. In that case, it is still possible to go through all hyperedges and find a combination of pairwise disjoint hyperedges that leads to maximum cardinality. For $N=3$ parties, the problem can also be solved via Network Flow. For larger \emph{N}, all combinations of hyperedges are to be considered and the optimal one is chosen. 
Details about the implementation can be found in App. \ref{sec:B}. 

In addition to finding a maximum matching, weights can be added to the edges and optimized in a second step. 
In our scenario, the weights will be a function of the number of rounds that the relevant qubits have spent in the memories. 
Note, that the main goal does not change and a maximum matching is to be found first. If there is more than one such maximum matching, the one with optimized weights is chosen. This leads to the \textsc{Maximum N-dimensional matching with weights}. Here, it is always necessary to go through all solutions and choose the optimal matching. 
How the weights are set is defined in Sec. \ref{sec:Strategies}. The implementation details are given in App. \ref{sec:B}.

\subsection{The router rate}
\label{sec:RepeaterRate}

\begin{figure}
    \centering
    \includegraphics[scale=0.5]{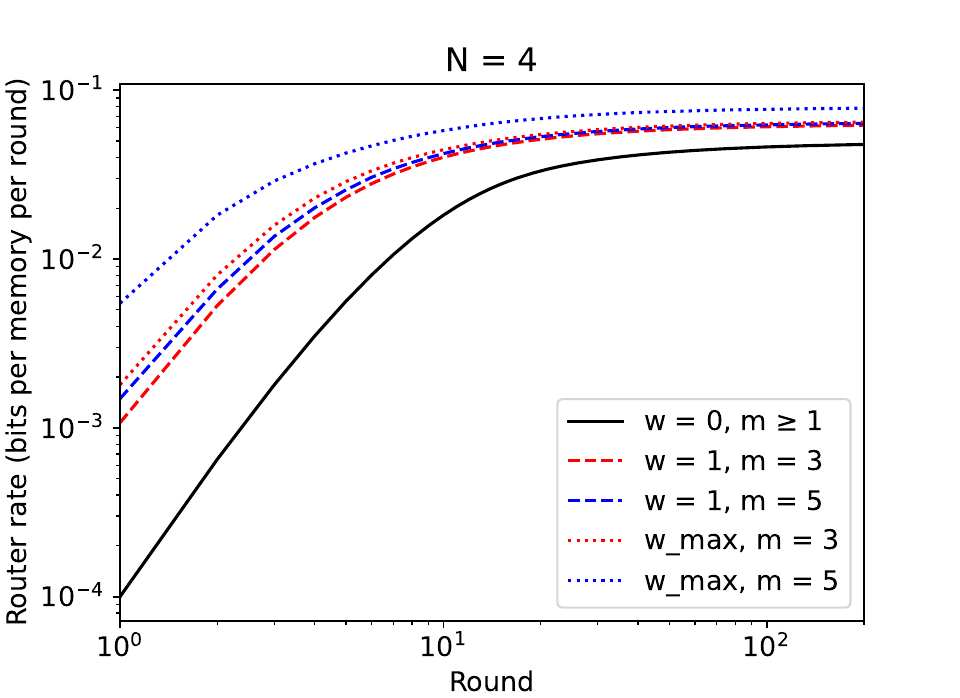}
    \caption{Comparison of the router rate for different 4-partite networks with varying numbers of memories \emph{m} per party and different maximal connection lengths \emph{w}. The transmittivity is chosen to be $\eta=0.1$. }
    \label{fig:4partite}
\end{figure}

In this subsection, we will analyze the router rate 
\begin{align}
    R(s_c) = \frac{1}{s_c} \sum_{s=1}^{s_c} \frac{\langle l \rangle (s)}{m} \label{eq:RepRate},
\end{align}
as defined in Sec. \ref{sec:QR}. Following \cite{Multiplexing}, we find analytical expressions for the generalized rate in a quantum router connecting more than two parties. 
We start by defining the configuration of the memories belonging to each party via the vectors $\Vec{a}, \Vec{b_1}, \Vec{b_2}, ...$; each of length $m$. Every entry in a vector represents one memory from the set of memories per party and can be either 0 (empty memory) or 1 (filled memory). In the example given in Fig. \ref{fig:matching_w_dep}, the vectors describing the memory configurations are: $\Vec{a} = (1, 0, 1, 0), \Vec{b_1} = (1, 1, 0, 1)$ and $\Vec{b_2} = (0, 0, 1, 1)$. From this, the total memory configuration $\mathcal{C} = \left(\Vec{a}_1, \Vec{b}_1, \hdots, \Vec{b}_{N-1} \right)$ follows; a vector that is given by the concatenation of the memory configuration of each party.  
Again, each entry $c_i \in \{0,1\}$ in $\mathcal{C}$ represents one memory. The value is set to 0 for empty memories and it is set to 1 when the memory is filled.   
As the maximal connection length $w$ limits the number of GHZ measurements, the set of configurations of $m$ memories per party leading to $l$ GHZ measurements is further denoted as $\mathcal{H}_w^m(l)$. Regarding Fig. \ref{fig:matching_w_dep} as an example, then $\left(\Vec{a}, \Vec{b_1}, \Vec{b_2}\right) \in \mathcal{H}_3^4(2)$, $\left(\Vec{a}, \Vec{b_1}, \Vec{b_2}\right) \in \mathcal{H}_2^4(2)$, $\left(\Vec{a}, \Vec{b_1}, \Vec{b_2}\right) \in \mathcal{H}_1^4(1)$, and $\left(\Vec{a}, \Vec{b_1}, \Vec{b_2}\right) \in \mathcal{H}_0^4(0)$ resulting in no matching. 
However, this configuration does e.g. not allow $l=2$ and $w=1$, i.e. $\left(\Vec{a}, \Vec{b_1}, \Vec{b_2}\right) \notin \mathcal{H}_1^4(2)$, 

With these notations, the transition between two memory configurations can be denoted via the storage map $\sigma_l: \mathcal{H}_w^m(0)\rightarrow \mathcal{H}_w^m(l)$ and the measurement map $\mu_l: \mathcal{H}_w^m(l)\rightarrow \mathcal{H}_w^m(0) $, see also Fig. \ref{fig:MultiRound}. 
The former contains the transition from configuration $\mathcal{C}$ to $\mathcal{C'}$ given by the probability $\eta $ of successfully sending each of the qubits through the quantum channel:
\begin{align}
    \text{Prob}\left[\sigma_l(\mathcal{C}) = \mathcal{C'}\right] &= \text{Prob}\left[\mathcal{C}' | \mathcal{C}\hspace{0.1cm}\right]  \nonumber \\
    &= \prod_{i=1}^{N\cdot m} \text{Prob} \left[c_i ' | c_i\right] \label{eq:1} \nonumber \\
    &= \prod_{i=1}^m \text{Prob}[a_i ' | a_i] \prod_{j=1}^{N-1} \text{Prob}\left[b_{j,i}' | b_{j,i}\right]
\end{align}
The transition probability between an initial configuration $\mathcal{C}$ and the configuration $\mathcal{C'}$ is calculated memory-wise for each configuration entry $c_i$ and $c'_i$ with $c_i, c_i' \in \{0,1\}$ and $i\in \{1, 2, \hdots, N\cdot m \}$
(i.e. $a_i$ for Alice's memories and $b_{j,i}$ with $i\in \{1, \hdots, m \}$ for the memories of the $j$ Bobs)
by 
\begin{align}
    \text{Prob}\left[c_i '|c_i \right] =& \hspace{0.1cm}(1-\eta)(1-c_i')(1-c_i) \nonumber \\
    &+\eta c_i'(1-c_i)+c_i'c_i  
\end{align}
The map $\mu_l$ describes transitions given by the GHZ measurements, i.e. it maps the memory configurations $\mathcal{C}'$ before the measurements to after the measurements ($\mathcal{C}''$). The choice of the memories included in such a measurement is made based on the underlying matching problem explained in the previous section. Combining both maps, the probability for an initial memory configuration $\mathcal{C}$ to end in the final configuration $\mathcal{C}''$ is given by:
\begin{multline}
    \text{Prob} \left[ \mu_l \circ \sigma_l \left(\mathcal{C} \right) = \mathcal{C}''\right] = \sum_{\mathcal{C'} \in \mathcal{H}_w^m(l)} \delta_{\mu_l(\mathcal{C}'), \mathcal{C}''} \cdot \\ \text{Prob} \left[ \sigma_l(\mathcal{C}) = \mathcal{C}'\right] \label{eq:2}
\end{multline}
where $\delta$ denotes the Kronecker delta that is 1 iff $\mu_l(\mathcal{C}') = \mathcal{C}''$ and 0 otherwise. 
\\
\indent
Remark: The main difference in the calculation of the router rate for $N>2$ parties (compared to $N=2$) lies in the determination of these two transition maps. The representation of a configuration $\mathcal{C}$ now includes the concatenation of \emph{N} vectors, instead of only two vectors. The verification of whether a configuration is in $\mathcal{H}_w^m(l)$ must be adjusted accordingly. 
\\
\indent 
Using these generalized transition matrices, and given that all memories are empty at the beginning of the protocol, the router rate can be calculated analytically analogously to \cite{Multiplexing}. 
That is done by computing each possible configuration $\mathcal{C}''_s$ at the end of one round $s$ (which equals the configuration $\mathcal{C}_{s+1}$ at the beginning of the next round) iteratively by knowing the transition probabilities given by Eq. (\ref{eq:2}) and the final configuration from the end of the previous round denoted by $\mathcal{C}''_{s-1} = \mathcal{C}_s$:
\begin{multline}
    \text{Prob} [\mathcal{C}''_s](s) = \sum_{\mathcal{C}_s \in \mathcal{H}_w^m(0)} \sum_{l=0}^m \text{Prob} \left[ \mu_l \circ \sigma_l (\mathcal{C}_s) = \mathcal{C}''_s\right] \cdot \\ \text{Prob} \left[ \mathcal{C}_s\right] (s)  
    = \text{Prob}[\mathcal{C}_{s+1}]
\end{multline}
Given this probability for any configuration at the beginning of a round ($\text{Prob}[\mathcal{C}](s)$), we calculate the probability of having $l$ GHZ measurements:
\begin{multline}
    \text{Prob} \left[ \Lambda=l \right](s) = \sum_{\mathcal{C}' \in \mathcal{H}_w^m(l)} \text{Prob} \left[ \mathcal{C}'\right](s)  \\
    = \sum_{\mathcal{C}' \in \mathcal{H}_w^m(l)} \sum_{\mathcal{C} \in \mathcal{H}_w^m(0)} \text{Prob} [\sigma_l(\mathcal{C}) = \mathcal{C'}] \text{Prob} [\mathcal{C}](s)
\end{multline}
Here, $\Lambda$ denotes a random variable that can take values from $0,1,...,m$ representing the number of performed GHZ measurements. In the case of considering not only deterministic GHZ measurements but also probabilistic measurements, the probability that $l$ GHZ measurements are performed successfully is given by 
\begin{multline}
    \text{Prob} \left[ \Sigma=l \right](s) = \sum_{i=l}^m \left( \begin{array}{llll}
i\\
l\\
\end{array} \right) \text{Prob}\left[ \Lambda = i \right] (s) \cdot \\ 
P_{GHZ}^l \cdot (1-P_{GHZ})^{i-l}
\end{multline}
with $P_{GHZ}$ being the success probability of a GHZ measurement. $\Sigma$ is also a random variable taking values $0,1,...,m$ defining the number of successful GHZ measurements. Then, the average number of successful measurements is given by: 
\begin{align}
    \langle l \rangle (s) = \sum_{l=0}^m l \cdot \text{Prob} \left[\Sigma=l\right] (s) \label{eq:11}
\end{align}
Finally, by inserting this in Eq. (\ref{eq:RepRate}) the router rate can be calculated. 
\\
\indent 
Note, that the two matrices describing the transitions of the memory configurations are of dimension $2^{m_{tot}} \times 2^{m_{tot}}$ with $m_{tot} = Nm$ being the total number of memories. This limits the network size, for which the router rate can be calculated analytically. For $m_{tot} > 12$, the runtime becomes infeasible and makes analytical calculations impracticable. 
All results considering larger networks are based on numerical simulations performing the presented protocol and extracting the average number of GHZ measurements per round from 50,000 repetitions. The simulation is further described in App. \ref{sec:A}.   
\\
\\
\indent
The router rate for different network sizes is calculated for the following parameters: the probability of the successful transmission of a qubit is $\eta=0.1$, which is equivalent to a distance between each party and the central router of $d = -10/\alpha \cdot \log_{10} \eta = 50$ km for a fiber attenuation coefficient of $\alpha = 0.2$. This value describes a commercial optical fiber used with light at a wavelength of 1550 nm. The GHZ measurement is assumed to be perfect.
In addition, it is assumed that all memories are initially empty and qubits remain in the memories until they are selected for a GHZ measurement by matching. The impact of premature removal of qubits from storage will be discussed later. Fig. \ref{fig:4partite} shows the router rates for a 4-partite network with various numbers of memories per party. 
For each choice of $m$, a maximal connection length of $w=0$, $w=1$, and $w=m-1$ is chosen. 
\par
\begin{figure}
    \centering
    \includegraphics[scale=0.5]{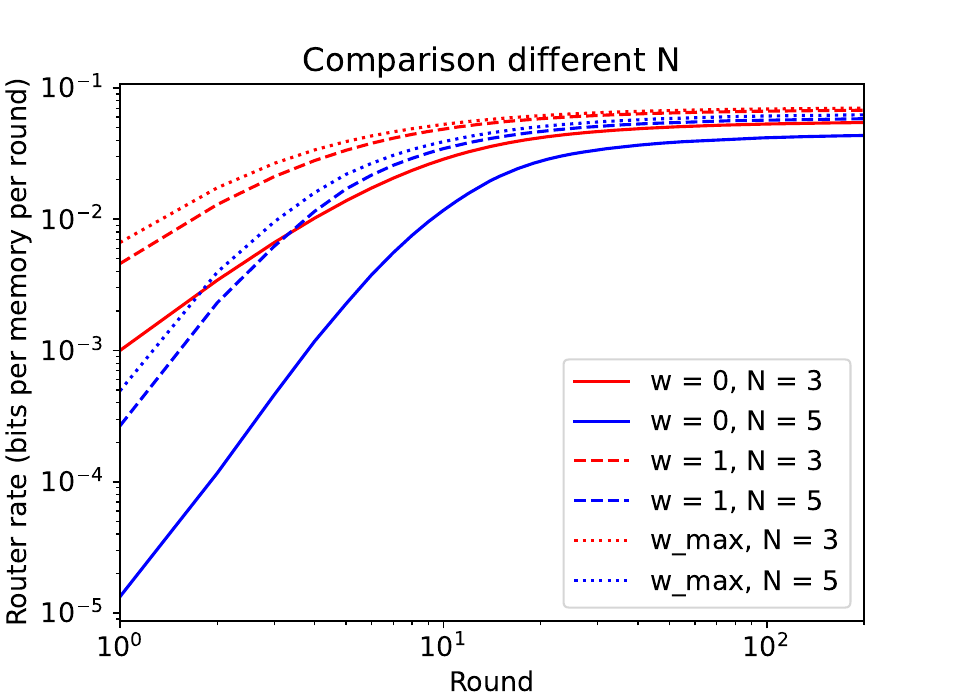}
    \caption{Comparison of the router rate for different network sizes where each party has $m=3$ memories. Parameters: $w=0, 1, 2$, $\eta=0.1$.}
    \label{fig:4memories} 
\end{figure}

 
In the case of $w=1$, only a small increase in the router rate with increasing \emph{m} can be seen in Fig. \ref{fig:4partite}. The biggest difference in the router rate for increasing \emph{m} is achieved with full-range multiplexing. The plot shows that for small $m$ a large advantage can already be gained by using finite-range multiplexing, e.g. $w=1$. As the number of memories per party increases, the increase of the router rate with the connection length \emph{w} becomes larger. 
\\ 
\\
\indent
The relationship between the number of communicating parties and the router rate is shown in Fig. \ref{fig:4memories} for $N=3$ and $N=5$, with a fixed memory number of $m=3$.
Again, a significant improvement in the router rate can already be achieved by considering finite-range multiplexing with $w=1$. The simulations also show, that for larger \emph{N} the router rate decreases significantly but the general behavior of each graph does not change. The fast decrease of the router rate for larger $N$ points out the importance of the usage of memories in a quantum router. The effect of decoherence that comes along with the memories is analyzed in the following.

%------------------------ CKA -------------------------------------------

\section{Secret key rate} 
\label{sec:CKA}

When the entangled states are distributed between all parties, these states can be used for further applications such as conference key agreement \cite{CKA}. Given that a GHZ state is shared between all parties, every party measures his/ her local part of the GHZ state either in the $Z$-basis (key generation round) or in the $X$-basis (test round). Additionally, the usual post-processing steps for CKA are performed as given in \cite{CKA}. 
Using some of the measurement outcomes for the classical parameter estimation, the quantum bit error rates $Q_X$ and $Q_{AB_i}$ can be estimated. 
In particular, $Q_{AB_i}$ is the probability that A and B$_i$ have different outcomes in the $Z$-basis measurement. $Q_X$ is the error rate for the measurements in the $X$-basis with respect to $|GHZ \rangle$. Explicitly, these QBERs are defined by:
\begin{align}
    Q_X  &= \frac{1- \langle X^{\otimes N} \rangle}{2},  \label{eq:Qx} 
\end{align}
\begin{align}
    Q_{AB_i} &= \text{Prob}\left( Z_A \neq Z_{B_i} \right) \nonumber \\
    &= \frac{1- \langle Z_A Z_{B_i} \rangle}{2}.  \label{eq:Qabi}
\end{align}
The QBERs in general depend on the number of storage rounds of all parties (for the simulations see App. \ref{sec:A} for details).

The asymptotic secret fraction $r_\infty$, which is a central figure of merit, is also calculated. It defines the fraction of secret bits that can be extracted from the measured qubits.
For the generalized $N$-BB84 protocol \cite{CKA, NBB84} the  asymptotic secret fraction is as follows:  
\begin{align}
    \label{eq:rinfty}
    r_\infty \left( Q^{tot}_X, Q^{tot}_{AB_i} \right) &= 
    \max \left[ 0, 1- h(Q^{tot}_X) - \max_{1\leq i \leq N-1} h(Q^{tot}_{AB_i}) \right]   
\end{align}
with the binary Shannon entropy $h(p) = -p \log_2(p) - (1-p) \log_2 (1-p)$. 
The total QBER $Q^{tot} \in \{Q_X^{tot}, Q^{tot}_{AB_i} \}$ is given by the total number of measurements per round multiplied by the average QBER in each round divided by the total number of measurements summed over the whole number of rounds up to $s_c$:  
\begin{widetext} 
\begin{align}
\label{eq:totalQ}
    Q^{tot}(s_c) = \frac{\sum_{s=1}^{s_c} \langle l \rangle (s) \sum_{\delta_{a}}^{s} \hdots \sum_{\delta_{b_{N-1}}}^{s} Q(\delta_{a}, \hdots, \delta_{b_{N-1}}) \text{Prob}\left[ \delta_{a}  \right] \hdots \text{Prob}\left[ \delta_{b_{N-1}}  \right] (s)}{\sum_{s=1}^{s_c} \langle l \rangle (s)}
\end{align}
\end{widetext}
Here, $\text{Prob}[\delta_i]$ with $i \in \{ a, b_1, \hdots, b_{N-1} \}$ is the fraction of qubits of party $i$ that are used in a GHZ measurement that has experienced decoherence for a certain number of rounds $\delta$.
The term $Q(\delta_{a}, \hdots, \delta_{b_{N-1}})$ is the theoretical QBER for the given storage times, which depend on all the fidelities of the stored qubits $a, b_{1}, \hdots, b_{N-1}$. The probability of white noise $p(\delta_i) = e^{-\delta_i/\tau}$ depends on all the storage rounds of all parties $\delta_{a}, \delta_{b_1}, \hdots, \delta_{b_{N-1}}$.  
\par
Finally, the secret key rate $K(s_c)$ can be calculated as the product of the asymptotic secret fraction from Eq. (\ref{eq:rinfty}) and the router rate from Eq. (\ref{eq:RepRate}):
\begin{align}
\label{eq:KeyRate}
    K(s_c) = r_\infty \left(Q^{tot}_X(s_c), Q^{tot}_{AB_i}(s_c) \right) \cdot R(s_c)
\end{align}
In App. \ref{sec:Tripartite}, an analytic calculation of the shared state after the GHZ measurement and the resulting quantum bit error rates is presented for the tripartite network. Using these results, we analyze different matching strategies with the aim of finding the one that optimizes the secret key rate. This comparison is presented in an exemplary way in the tripartite router setup in the following subsections. 

\subsection{Strategies for storage and measurement}

\label{sec:Strategies}

In \cite{Multiplexing}, various strategies for how to choose memories for the BSMs were analyzed, with the aim of maximizing the secret key rate. 
Here, we generalize the strategies to $N$ parties.  
To integrate strategies into the matching process defined in Sec. \ref{sec:match}, weights are associated with the edges, taking into account the number of storage rounds $\delta$ of the qubits that are included in a GHZ measurement. Two different ways are chosen to calculate the weights:
\begin{enumerate}
    \item The weight $W_1$ of a hyperedge is defined by the absolute values of the difference in the number of storage rounds per two qubits, summed over all relevant bipartite edges within the hyperedge (see construction of the graph as e.g. in Figure \ref{fig:matching_w_dep}). For the case $N=3$ this reads:
    \begin{align}
        \label{eq:w1}
        W_1 &= |\delta_{b_1} - \delta_{a}| + |\delta_{a} - \delta_{b_2}|
    \end{align}
    \item The weight $W_2$ is defined by the sum of the number of storage rounds of each qubit contained in the hyperedge. Explicitly for the case $N=3$:
    \begin{align}
    \label{eq:w2}
        W_2 &= \delta_{b_1} + \delta_{a} + \delta_{b_2}
    \end{align}
\end{enumerate}
Based on this, we compare the following strategies for the choice of the maximum matching among several possible ones: 
\begin{enumerate}
    \item[S0:] The first maximum matching found is chosen, independently of the number of storage rounds. 
    \item[S1:] Taking the number of storage rounds of the qubits in the memories into account we 
    \begin{enumerate}
        \item[a.] minimize over the sum of weights $W_1$ (defined in Eq. (\ref{eq:w1})) of all hyperedges. 
        \item[b.] maximize over the sum of weights $W_1$ (defined in Eq. (\ref{eq:w1})) of all hyperedges.   
    \end{enumerate}
    \item[S2:] Taking the number of storage rounds of the qubits in the memories into account we minimize the sum of weights $W_2$ (defined in Eq. (\ref{eq:w2})) of all hyperedges.  
\end{enumerate}
Strategy S1 a. produces states with the highest correlations by preferably choosing qubits with a minimal difference in the number of storage rounds, while qubits with large differences in the number of storage rounds are left over. As a consequence, older qubits are chosen less often than newer qubits. 
On the opposite, older qubits are chosen earlier by combining them with the newest qubits in Strategy S1 b. In strategy S2, we focus on a maximal fidelity by connecting qubits with the lowest number of storage rounds. Therefore, new qubits are measured as soon as possible. 
For example, with a memory allocation of $\delta_{a} = 0$ or $\delta_{a} = 1$, $\delta_{b_1} = 3$, and $\delta_{b_2} = 3$, strategy S1 a. would select the qubits with $\delta_{a} = 1$, $\delta_{b_1} = 3$, and $\delta_{b_2} = 3$, while Strategy S2 would select the qubits with $\delta_{a} = 0$, $\delta_{b_1} = 3$, and $\delta_{b_2} = 3$ measuring newer qubits first.

Note, that a maximization of the number of storage rounds (leading to a maximization of strategy S2) is not considered, as this increases the QBER and therefore leads to a decreasing secret key rate.  
\begin{figure}
    \centering
    \includegraphics[scale=0.5]{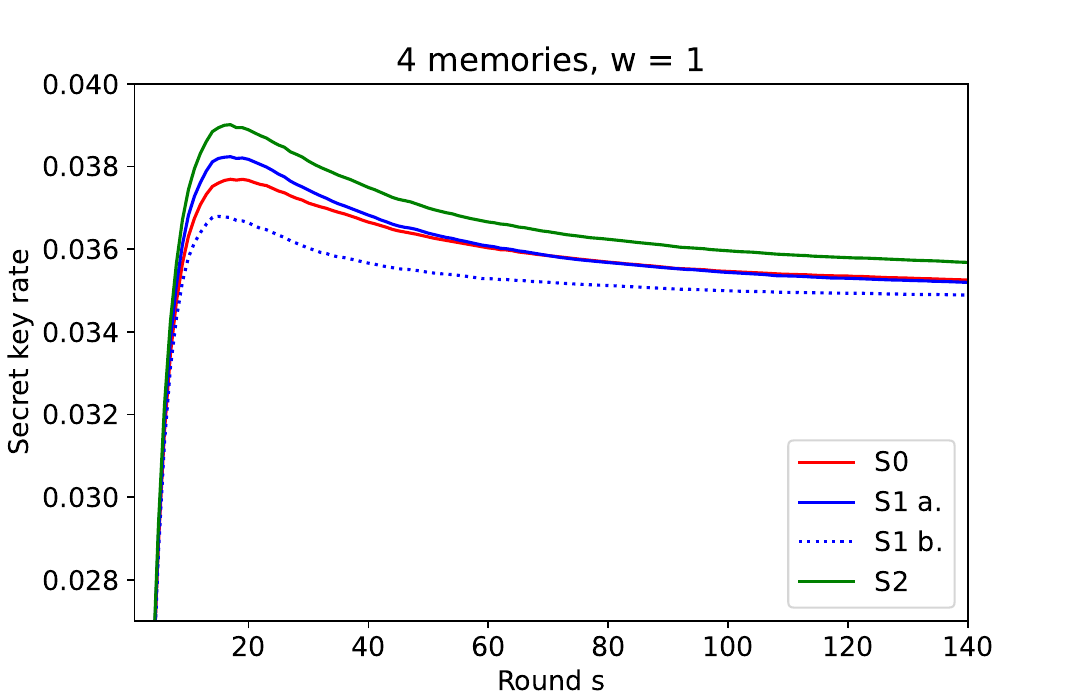}
    \caption{Comparison of the secret key rate for the different strategies as described in Sec. \ref{sec:Strategies}. In the tripartite network, each party has 4 memories and the maximal connection length is set to $w=1$. Parameters: $\eta=0.1$, $\tau = 100$, 50,000 samples. }
    \label{fig:enter-label}
\end{figure}

Fig. \ref{fig:enter-label} shows the secret key rate computed in a tripartite network with 4 memories per party. The maximal connection length is set to finite-range multiplexing ($w=1$). The calculations are based on simulations with 50,000 samples to get the average number of GHZ measurements per round ($\langle l \rangle$) and the probability for the number of storage rounds ($\text{Prob}[\delta]$). Details of the simulation are given in App. \ref{sec:A}. The transmittivity is set to $\eta=0.1$ and $\tau = 100$ is the decoherence parameter defined in Eq. (\ref{eq:deco}).  
It turns out that considering the highest fidelities (i.e. the sum over the number of storage rounds has to be minimized, see strategy S2) leads to the highest secret key rate. Minimizing the difference in the number of rounds (see strategy S1 a.) leads to better results than choosing the first matching found ( strategy S0) when performing only a few number of rounds. In the long-term, that changes and strategy S0 performs slightly better than strategy S1 a. 
Nevertheless, strategy S2 is the only strategy of practical use independent of the number of protocol rounds being performed.   
\par
Note that a maximal secret key rate is reached after about 16 rounds as seen in Fig. \ref{fig:enter-label}. Exceeding this number of rounds, the secret key rate decreases due to the decoherence that the qubits experience in the memories. To reduce this effect, additional cutoffs will be introduced in a next step. Note, that strategy S2 already leads to the idea of cutoffs, since older qubits are chosen with lower preference compared to new qubits.

\subsection{Strategy: emptying the memories after cutoffs}

We now modify the protocol presented in Sec. \ref{sec:protocol} by introducing cutoffs $s_{cutoff}$ at the end of each round: older qubits are removed from the memories such that they can be refilled in upcoming rounds. The previously defined strategy S2 of matching the newest qubits first is kept here and cutoffs are additionally considered. 

\begin{figure}
    \centering
    \includegraphics[scale=0.41]{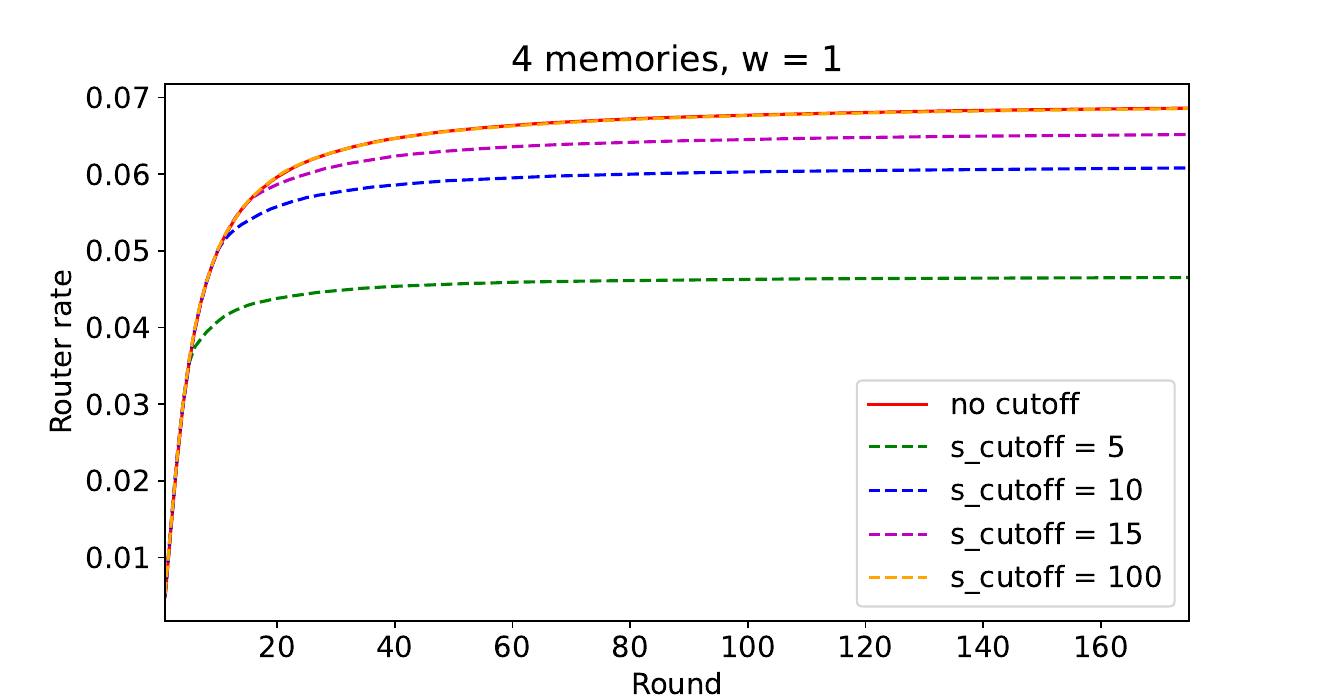} \\ (a) \\
    \includegraphics[scale=0.41]{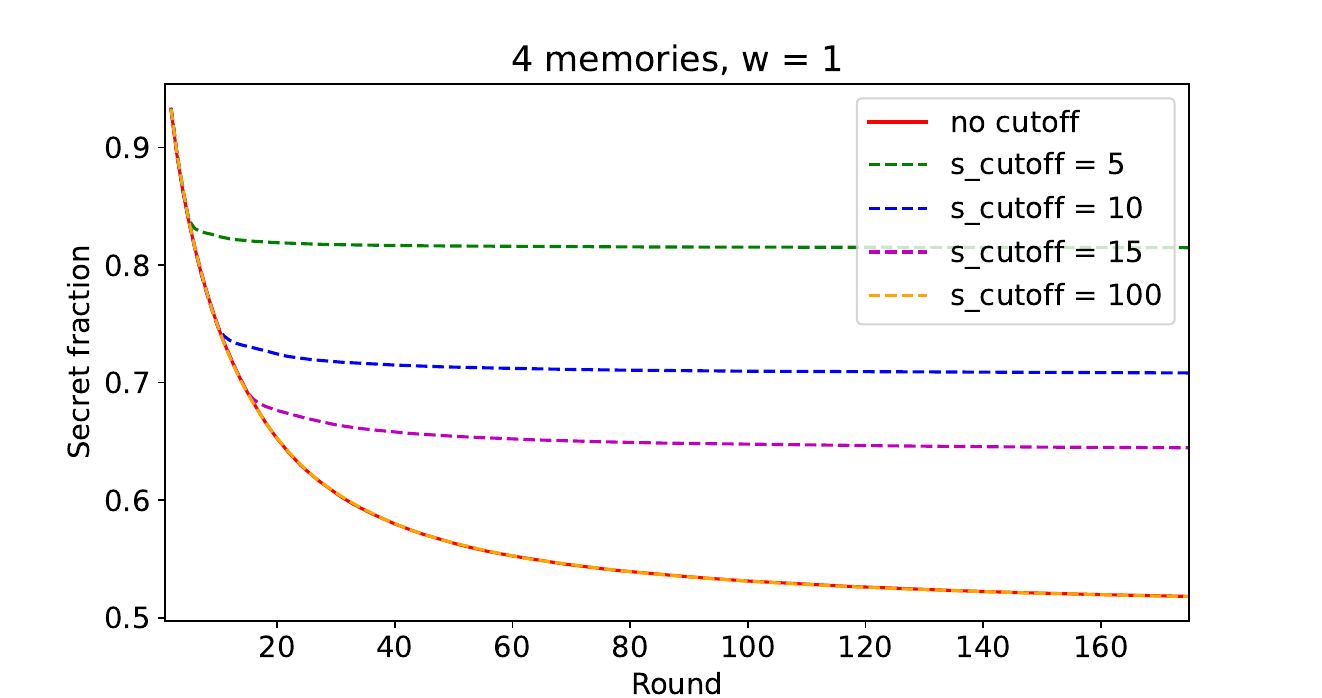} \\ (b) \\
    \includegraphics[scale=0.41]{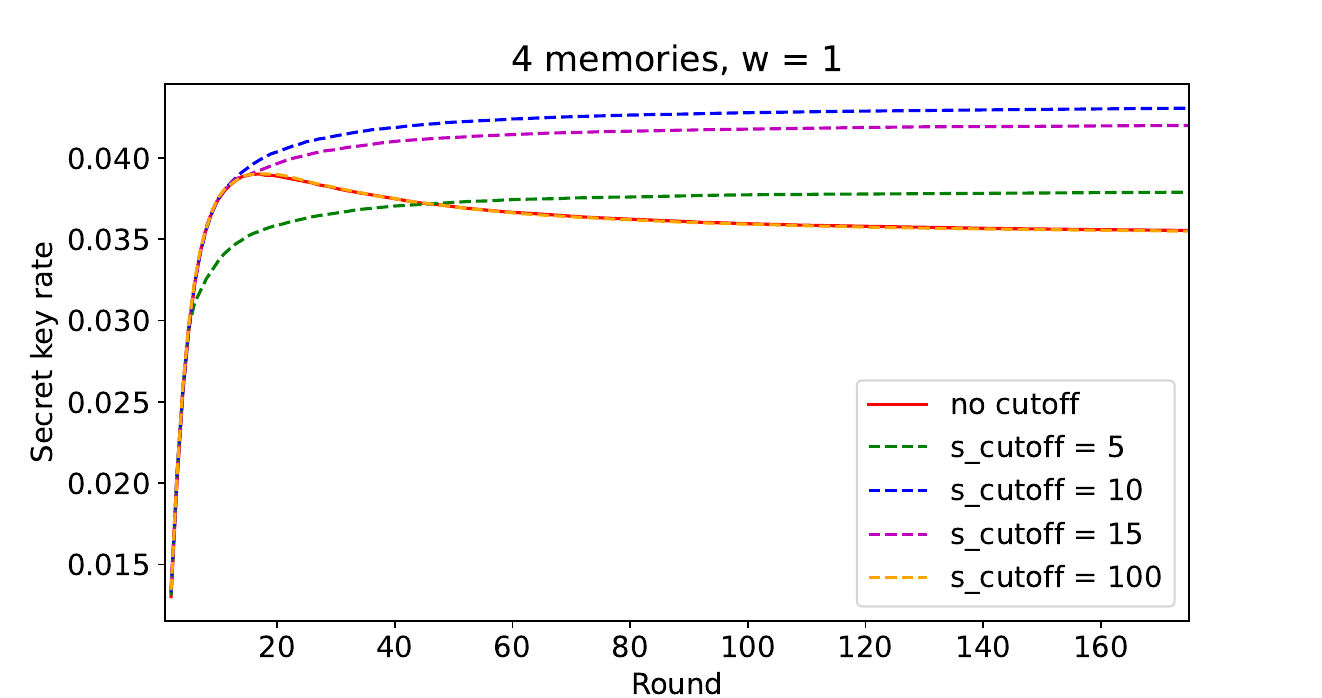} \\ (c)
    \caption{Effect of number of cutoff rounds on rates in a tripartite network with $m=4$ memories per party and finite-range multiplexing $w=1$. In (a) the router rate is shown for different cutoffs. (b) shows the secret fraction and (c) the secret key rate for the same setup. Parameters: $\eta=0.1$, $\tau = 100$, 50,000 samples.}
    \label{fig:Overview_with_cutoffs}
\end{figure}

The secret key rate plotted in Fig. \ref{fig:enter-label} shows the competing behavior of the router rate and the asymptotic secret fraction. Due to the exponential decay of the fidelities that the qubits experience while being stored a smaller cutoff leads to higher fidelity, which increases the asymptotic secret fraction. On the other side, the router rate decreases with smaller cutoffs, since a reduction in the cutoff comes along with a decreasing probability for memories to be filled. 
This can be seen in Fig. \ref{fig:Overview_with_cutoffs} (a) and Fig. \ref{fig:Overview_with_cutoffs} (b), where the router rate and the asymptotic secret fraction are plotted for different cutoffs. 
An optimal cutoff can be found where the two values multiply in a way that the secret key rate does not decrease with the number of rounds.  
\\
\indent
For our setup, such an optimal cutoff is achieved at $s_{cutoff} = 10$ rounds (seen by the simulations which were also done for $s_{cutoff} = 9$ and $s_{cutoff} = 11$), which leads to the maximum secret key rate, as seen in Fig. \ref{fig:Overview_with_cutoffs} (c). This cutoff can either be deduced from Fig. \ref{fig:enter-label}, where the secret key rate reaches roughly its maximum at about 16 rounds, or it can be argued mathematically via Eq. (\ref{eq:rinfty}) for the secret fraction. Assuming that $Q_X = Q_{AB_i}$, none of the QBERs should be larger than 0.11 since otherwise $r_\infty = 0$. To achieve $Q_X \geq 0.11$, the mean fidelity that the qubits need to have when included in a GHZ measurement can be calculated. Here, we assume that $F_A = 1$ and $F_{B_1} = F_{B_2}$ since one qubit is always new in the case of maximizing $l$. With the given parameters, $Q_X \geq 0.11$ is reached, when the qubits are maximally stored for $s_{cutoff} = 12$ rounds. This is only an approximation since several assumptions are made here.

\section{Conclusion}
\label{sec:Conclusion}

Quantum routers form a main ingredient when building quantum networks as they increase the communication distance between end users. By including quantum memories and multiplexing, the router rate (i.e. entanglement generation) and the secret key rate can be improved significantly. 
In this work, we presented a generalization of the bipartite repeater to the quantum router in an $N$-partite star graph with the quantum router being the central node. 
In contrast to previous work, we integrated quantum memories into the quantum router and considered multiplexing. 
We described the underlying graph-theoretical matching problem and implemented different algorithms for the \textsc{maximum $N$-dimensional matching (with weights)} to reduce the runtime as much as possible. 
\par 
Furthermore, we analyzed the router rate in star networks up to a network size of 5 parties with 4 memories each. 
The resulting plots show that even for finite-range multiplexing large improvements in the router rate are achieved. 
In a second step, we considered conference key agreement and studied noise effects on the quantum memories. Due to decoherence, the asymptotic secret fraction, and therefore the secret key rate, decreases when qubits are stored longer. To minimize this effect, we have investigated different storage and measurement strategies for the GHZ measurement. It turned out that it is best to connect the qubits with the shortest number of storage rounds (highest fidelities), i.e. to minimize the sum of the number of storage rounds in each matching. This strategy outperforms all other investigated strategies.   
\par
We have further modified the protocol removing older qubits from the memory after a certain number of rounds since they can not contribute to the secret key rate. 
Combining the optimal matching strategy with the optimal number of cutoff rounds leads to the overall highest secret key rate.

In future work the analysis of larger networks with more than one central router will be of interest. It should be investigated how the network structure influences the router rate and the secret key rate depending on multiplexing with different matching strategies. It might be also conceivable to extend the protocols and, for example, to take distillation into account. Regarding implementations, the influence of finite key effects might be integrated as well, following \cite{finite}.

\section{Acknowledgment}

We thank Dorothea Baumeister, Gláucia Murta, Jörg Rothe, and Nikolai Wyderka for valuable discussions. This work was funded by the Federal Ministry of Education and Research BMBF (Project QR.X).  

\bibliography{bibliography.bib}

\begin{thebibliography}{20}%
\makeatletter
\providecommand \@ifxundefined [1]{%
 \@ifx{#1\undefined}
}%
\providecommand \@ifnum [1]{%
 \ifnum #1\expandafter \@firstoftwo
 \else \expandafter \@secondoftwo
 \fi
}%
\providecommand \@ifx [1]{%
 \ifx #1\expandafter \@firstoftwo
 \else \expandafter \@secondoftwo
 \fi
}%
\providecommand \natexlab [1]{#1}%
\providecommand \enquote  [1]{``#1''}%
\providecommand \bibnamefont  [1]{#1}%
\providecommand \bibfnamefont [1]{#1}%
\providecommand \citenamefont [1]{#1}%
\providecommand \href@noop [0]{\@secondoftwo}%
\providecommand \href [0]{\begingroup \@sanitize@url \@href}%
\providecommand \@href[1]{\@@startlink{#1}\@@href}%
\providecommand \@@href[1]{\endgroup#1\@@endlink}%
\providecommand \@sanitize@url [0]{\catcode `\\12\catcode `\$12\catcode
  `\&12\catcode `\#12\catcode `\^12\catcode `\_12\catcode `\%12\relax}%
\providecommand \@@startlink[1]{}%
\providecommand \@@endlink[0]{}%
\providecommand \url  [0]{\begingroup\@sanitize@url \@url }%
\providecommand \@url [1]{\endgroup\@href {#1}{\urlprefix }}%
\providecommand \urlprefix  [0]{URL }%
\providecommand \Eprint [0]{\href }%
\providecommand \doibase [0]{https://doi.org/}%
\providecommand \selectlanguage [0]{\@gobble}%
\providecommand \bibinfo  [0]{\@secondoftwo}%
\providecommand \bibfield  [0]{\@secondoftwo}%
\providecommand \translation [1]{[#1]}%
\providecommand \BibitemOpen [0]{}%
\providecommand \bibitemStop [0]{}%
\providecommand \bibitemNoStop [0]{.\EOS\space}%
\providecommand \EOS [0]{\spacefactor3000\relax}%
\providecommand \BibitemShut  [1]{\csname bibitem#1\endcsname}%
\let\auto@bib@innerbib\@empty
%</preamble>
\bibitem [{\citenamefont {Bennett}\ and\ \citenamefont
  {Brassard}(1984)}]{BB84_Original}%
  \BibitemOpen
  \bibfield  {author} {\bibinfo {author} {\bibfnamefont {C.~H.}\ \bibnamefont
  {Bennett}}\ and\ \bibinfo {author} {\bibfnamefont {G.}~\bibnamefont
  {Brassard}},\ }\bibfield  {title} {\bibinfo {title} {Quantum cryptography:
  Public key distribution and coin tossing},\ }\href@noop {} {\bibfield
  {journal} {\bibinfo  {journal} {Proceedings of the IEEE International
  Conference on Computers, Systems and Signal Processing}\ }\textbf {\bibinfo
  {volume} {175}} (\bibinfo {year} {1984})}\BibitemShut {NoStop}%
\bibitem [{\citenamefont {Ekert}(1991)}]{Ekert}%
  \BibitemOpen
  \bibfield  {author} {\bibinfo {author} {\bibfnamefont {A.~K.}\ \bibnamefont
  {Ekert}},\ }\bibfield  {title} {\bibinfo {title} {Quantum cryptography based
  on bell's theorem},\ }\href {https://doi.org/10.1103/PhysRevLett.67.661}
  {\bibfield  {journal} {\bibinfo  {journal} {Phys. Rev. Lett.}\ }\textbf
  {\bibinfo {volume} {67}},\ \bibinfo {pages} {661} (\bibinfo {year}
  {1991})}\BibitemShut {NoStop}%
\bibitem [{\citenamefont {Murta}\ \emph {et~al.}(2020)\citenamefont {Murta},
  \citenamefont {Grasselli}, \citenamefont {Kampermann},\ and\ \citenamefont
  {Bru{\ss}}}]{CKA}%
  \BibitemOpen
  \bibfield  {author} {\bibinfo {author} {\bibfnamefont {G.}~\bibnamefont
  {Murta}}, \bibinfo {author} {\bibfnamefont {F.}~\bibnamefont {Grasselli}},
  \bibinfo {author} {\bibfnamefont {H.}~\bibnamefont {Kampermann}},\ and\
  \bibinfo {author} {\bibfnamefont {D.}~\bibnamefont {Bru{\ss}}},\ }\bibfield
  {title} {\bibinfo {title} {Quantum conference key agreement: A review},\
  }\href {https://doi.org/10.1002%2Fqute.202000025} {\bibfield  {journal}
  {\bibinfo  {journal} {Advanced Quantum Technologies}\ }\textbf {\bibinfo
  {volume} {3}},\ \bibinfo {pages} {2000025} (\bibinfo {year}
  {2020})}\BibitemShut {NoStop}%
\bibitem [{\citenamefont {Grasselli}\ \emph {et~al.}(2018)\citenamefont
  {Grasselli}, \citenamefont {Kampermann},\ and\ \citenamefont
  {Bru{\ss}}}]{NBB84}%
  \BibitemOpen
  \bibfield  {author} {\bibinfo {author} {\bibfnamefont {F.}~\bibnamefont
  {Grasselli}}, \bibinfo {author} {\bibfnamefont {H.}~\bibnamefont
  {Kampermann}},\ and\ \bibinfo {author} {\bibfnamefont {D.}~\bibnamefont
  {Bru{\ss}}},\ }\bibfield  {title} {\bibinfo {title} {Finite-key effects in
  multipartite quantum key distribution protocols},\ }\href
  {https://doi.org/10.1088/1367-2630/aaec34} {\bibfield  {journal} {\bibinfo
  {journal} {New Journal of Physics}\ }\textbf {\bibinfo {volume} {20}},\
  \bibinfo {pages} {113014} (\bibinfo {year} {2018})}\BibitemShut {NoStop}%
\bibitem [{\citenamefont {Scarani}\ \emph {et~al.}(2009)\citenamefont
  {Scarani}, \citenamefont {Bechmann-Pasquinucci}, \citenamefont {Cerf},
  \citenamefont {Du\ifmmode~\check{s}\else \v{s}\fi{}ek}, \citenamefont
  {L\"utkenhaus},\ and\ \citenamefont {Peev}}]{Repeater}%
  \BibitemOpen
  \bibfield  {author} {\bibinfo {author} {\bibfnamefont {V.}~\bibnamefont
  {Scarani}}, \bibinfo {author} {\bibfnamefont {H.}~\bibnamefont
  {Bechmann-Pasquinucci}}, \bibinfo {author} {\bibfnamefont {N.~J.}\
  \bibnamefont {Cerf}}, \bibinfo {author} {\bibfnamefont {M.}~\bibnamefont
  {Du\ifmmode~\check{s}\else \v{s}\fi{}ek}}, \bibinfo {author} {\bibfnamefont
  {N.}~\bibnamefont {L\"utkenhaus}},\ and\ \bibinfo {author} {\bibfnamefont
  {M.}~\bibnamefont {Peev}},\ }\bibfield  {title} {\bibinfo {title} {The
  security of practical quantum key distribution},\ }\href
  {https://doi.org/10.1103/RevModPhys.81.1301} {\bibfield  {journal} {\bibinfo
  {journal} {Rev. Mod. Phys.}\ }\textbf {\bibinfo {volume} {81}},\ \bibinfo
  {pages} {1301} (\bibinfo {year} {2009})}\BibitemShut {NoStop}%
\bibitem [{\citenamefont {Briegel}\ \emph {et~al.}(1998)\citenamefont
  {Briegel}, \citenamefont {D\"ur}, \citenamefont {Cirac},\ and\ \citenamefont
  {Zoller}}]{Repeater1}%
  \BibitemOpen
  \bibfield  {author} {\bibinfo {author} {\bibfnamefont {H.-J.}\ \bibnamefont
  {Briegel}}, \bibinfo {author} {\bibfnamefont {W.}~\bibnamefont {D\"ur}},
  \bibinfo {author} {\bibfnamefont {J.~I.}\ \bibnamefont {Cirac}},\ and\
  \bibinfo {author} {\bibfnamefont {P.}~\bibnamefont {Zoller}},\ }\bibfield
  {title} {\bibinfo {title} {Quantum repeaters: The role of imperfect local
  operations in quantum communication},\ }\href
  {https://doi.org/10.1103/PhysRevLett.81.5932} {\bibfield  {journal} {\bibinfo
   {journal} {Phys. Rev. Lett.}\ }\textbf {\bibinfo {volume} {81}},\ \bibinfo
  {pages} {5932} (\bibinfo {year} {1998})}\BibitemShut {NoStop}%
\bibitem [{\citenamefont {D\"ur}\ \emph {et~al.}(1999)\citenamefont {D\"ur},
  \citenamefont {Briegel}, \citenamefont {Cirac},\ and\ \citenamefont
  {Zoller}}]{QuantumRepeater2}%
  \BibitemOpen
  \bibfield  {author} {\bibinfo {author} {\bibfnamefont {W.}~\bibnamefont
  {D\"ur}}, \bibinfo {author} {\bibfnamefont {H.-J.}\ \bibnamefont {Briegel}},
  \bibinfo {author} {\bibfnamefont {J.~I.}\ \bibnamefont {Cirac}},\ and\
  \bibinfo {author} {\bibfnamefont {P.}~\bibnamefont {Zoller}},\ }\bibfield
  {title} {\bibinfo {title} {Quantum repeaters based on entanglement
  purification},\ }\href {https://doi.org/10.1103/PhysRevA.59.169} {\bibfield
  {journal} {\bibinfo  {journal} {Phys. Rev. A}\ }\textbf {\bibinfo {volume}
  {59}},\ \bibinfo {pages} {169} (\bibinfo {year} {1999})}\BibitemShut
  {NoStop}%
\bibitem [{\citenamefont {Li}\ \emph {et~al.}(2019)\citenamefont {Li},
  \citenamefont {Zhang}, \citenamefont {Yin}, \citenamefont {Liu},
  \citenamefont {Hu}, \citenamefont {Fang}, \citenamefont {Fei}, \citenamefont
  {Jiang}, \citenamefont {Zhang}, \citenamefont {Li} \emph
  {et~al.}}]{Exprepeater}%
  \BibitemOpen
  \bibfield  {author} {\bibinfo {author} {\bibfnamefont {Z.-D.}\ \bibnamefont
  {Li}}, \bibinfo {author} {\bibfnamefont {R.}~\bibnamefont {Zhang}}, \bibinfo
  {author} {\bibfnamefont {X.-F.}\ \bibnamefont {Yin}}, \bibinfo {author}
  {\bibfnamefont {L.-Z.}\ \bibnamefont {Liu}}, \bibinfo {author} {\bibfnamefont
  {Y.}~\bibnamefont {Hu}}, \bibinfo {author} {\bibfnamefont {Y.-Q.}\
  \bibnamefont {Fang}}, \bibinfo {author} {\bibfnamefont {Y.-Y.}\ \bibnamefont
  {Fei}}, \bibinfo {author} {\bibfnamefont {X.}~\bibnamefont {Jiang}}, \bibinfo
  {author} {\bibfnamefont {J.}~\bibnamefont {Zhang}}, \bibinfo {author}
  {\bibfnamefont {L.}~\bibnamefont {Li}}, \emph {et~al.},\ }\bibfield  {title}
  {\bibinfo {title} {Experimental quantum repeater without quantum memory},\
  }\href@noop {} {\bibfield  {journal} {\bibinfo  {journal} {Nature photonics}\
  }\textbf {\bibinfo {volume} {13}},\ \bibinfo {pages} {644} (\bibinfo {year}
  {2019})}\BibitemShut {NoStop}%
\bibitem [{\citenamefont {Abruzzo}\ \emph
  {et~al.}(2014{\natexlab{a}})\citenamefont {Abruzzo}, \citenamefont
  {Kampermann},\ and\ \citenamefont {Bru{\ss}}}]{MDIQKD}%
  \BibitemOpen
  \bibfield  {author} {\bibinfo {author} {\bibfnamefont {S.}~\bibnamefont
  {Abruzzo}}, \bibinfo {author} {\bibfnamefont {H.}~\bibnamefont
  {Kampermann}},\ and\ \bibinfo {author} {\bibfnamefont {D.}~\bibnamefont
  {Bru{\ss}}},\ }\bibfield  {title} {\bibinfo {title}
  {Measurement-device-independent quantum key distribution with quantum
  memories},\ }\href {https://doi.org/10.1103%2Fphysreva.89.012301} {\bibfield
  {journal} {\bibinfo  {journal} {Physical Review A}\ }\textbf {\bibinfo
  {volume} {89}},\ \bibinfo {pages} {012301} (\bibinfo {year}
  {2014}{\natexlab{a}})}\BibitemShut {NoStop}%
\bibitem [{\citenamefont {Collins}\ \emph {et~al.}(2007)\citenamefont
  {Collins}, \citenamefont {Jenkins}, \citenamefont {Kuzmich},\ and\
  \citenamefont {Kennedy}}]{Multi_Original}%
  \BibitemOpen
  \bibfield  {author} {\bibinfo {author} {\bibfnamefont {O.~A.}\ \bibnamefont
  {Collins}}, \bibinfo {author} {\bibfnamefont {S.~D.}\ \bibnamefont
  {Jenkins}}, \bibinfo {author} {\bibfnamefont {A.}~\bibnamefont {Kuzmich}},\
  and\ \bibinfo {author} {\bibfnamefont {T.~A.~B.}\ \bibnamefont {Kennedy}},\
  }\bibfield  {title} {\bibinfo {title} {Multiplexed memory-insensitive quantum
  repeaters},\ }\href {https://doi.org/10.1103/PhysRevLett.98.060502}
  {\bibfield  {journal} {\bibinfo  {journal} {Phys. Rev. Lett.}\ }\textbf
  {\bibinfo {volume} {98}},\ \bibinfo {pages} {060502} (\bibinfo {year}
  {2007})}\BibitemShut {NoStop}%
\bibitem [{\citenamefont {Abruzzo}\ \emph
  {et~al.}(2014{\natexlab{b}})\citenamefont {Abruzzo}, \citenamefont
  {Kampermann},\ and\ \citenamefont {Bru{\ss}}}]{Multiplexing}%
  \BibitemOpen
  \bibfield  {author} {\bibinfo {author} {\bibfnamefont {S.}~\bibnamefont
  {Abruzzo}}, \bibinfo {author} {\bibfnamefont {H.}~\bibnamefont
  {Kampermann}},\ and\ \bibinfo {author} {\bibfnamefont {D.}~\bibnamefont
  {Bru{\ss}}},\ }\bibfield  {title} {\bibinfo {title} {Finite-range
  multiplexing enhances quantum key distribution via quantum repeaters},\
  }\href {https://doi.org/10.1103\%2Fphysreva.89.012303} {\bibfield  {journal}
  {\bibinfo  {journal} {Physical Review A}\ }\textbf {\bibinfo {volume} {89}},\
  \bibinfo {pages} {012303} (\bibinfo {year} {2014}{\natexlab{b}})}\BibitemShut
  {NoStop}%
\bibitem [{\citenamefont {Lo}\ \emph {et~al.}(2012)\citenamefont {Lo},
  \citenamefont {Curty},\ and\ \citenamefont {Qi}}]{MDIQKDOriginal}%
  \BibitemOpen
  \bibfield  {author} {\bibinfo {author} {\bibfnamefont {H.-K.}\ \bibnamefont
  {Lo}}, \bibinfo {author} {\bibfnamefont {M.}~\bibnamefont {Curty}},\ and\
  \bibinfo {author} {\bibfnamefont {B.}~\bibnamefont {Qi}},\ }\bibfield
  {title} {\bibinfo {title} {Measurement-device-independent quantum key
  distribution},\ }\href {https://doi.org/10.1103/PhysRevLett.108.130503}
  {\bibfield  {journal} {\bibinfo  {journal} {Phys. Rev. Lett.}\ }\textbf
  {\bibinfo {volume} {108}},\ \bibinfo {pages} {130503} (\bibinfo {year}
  {2012})}\BibitemShut {NoStop}%
\bibitem [{\citenamefont {Epping}\ \emph {et~al.}(2016)\citenamefont {Epping},
  \citenamefont {Kampermann},\ and\ \citenamefont {Bruß}}]{Epping}%
  \BibitemOpen
  \bibfield  {author} {\bibinfo {author} {\bibfnamefont {M.}~\bibnamefont
  {Epping}}, \bibinfo {author} {\bibfnamefont {H.}~\bibnamefont {Kampermann}},\
  and\ \bibinfo {author} {\bibfnamefont {D.}~\bibnamefont {Bruß}},\ }\bibfield
   {title} {\bibinfo {title} {Large-scale quantum networks based on graphs},\
  }\href {https://doi.org/10.1088/1367-2630/18/5/053036} {\bibfield  {journal}
  {\bibinfo  {journal} {New Journal of Physics}\ }\textbf {\bibinfo {volume}
  {18}},\ \bibinfo {pages} {053036} (\bibinfo {year} {2016})}\BibitemShut
  {NoStop}%
\bibitem [{\citenamefont {Coopmans}\ \emph {et~al.}(2021)\citenamefont
  {Coopmans}, \citenamefont {Knegjens}, \citenamefont {Dahlberg}, \citenamefont
  {Maier}, \citenamefont {Nijsten}, \citenamefont {de~Oliveira~Filho},
  \citenamefont {Papendrecht}, \citenamefont {Rabbie}, \citenamefont
  {Rozpędek}, \citenamefont {Skrzypczyk}, \citenamefont {Wubben},
  \citenamefont {de~Jong}, \citenamefont {Podareanu}, \citenamefont
  {Torres-Knoop}, \citenamefont {Elkouss},\ and\ \citenamefont
  {Wehner}}]{Coopmans_2021}%
  \BibitemOpen
  \bibfield  {author} {\bibinfo {author} {\bibfnamefont {T.}~\bibnamefont
  {Coopmans}}, \bibinfo {author} {\bibfnamefont {R.}~\bibnamefont {Knegjens}},
  \bibinfo {author} {\bibfnamefont {A.}~\bibnamefont {Dahlberg}}, \bibinfo
  {author} {\bibfnamefont {D.}~\bibnamefont {Maier}}, \bibinfo {author}
  {\bibfnamefont {L.}~\bibnamefont {Nijsten}}, \bibinfo {author} {\bibfnamefont
  {J.}~\bibnamefont {de~Oliveira~Filho}}, \bibinfo {author} {\bibfnamefont
  {M.}~\bibnamefont {Papendrecht}}, \bibinfo {author} {\bibfnamefont
  {J.}~\bibnamefont {Rabbie}}, \bibinfo {author} {\bibfnamefont
  {F.}~\bibnamefont {Rozpędek}}, \bibinfo {author} {\bibfnamefont
  {M.}~\bibnamefont {Skrzypczyk}}, \bibinfo {author} {\bibfnamefont
  {L.}~\bibnamefont {Wubben}}, \bibinfo {author} {\bibfnamefont
  {W.}~\bibnamefont {de~Jong}}, \bibinfo {author} {\bibfnamefont
  {D.}~\bibnamefont {Podareanu}}, \bibinfo {author} {\bibfnamefont
  {A.}~\bibnamefont {Torres-Knoop}}, \bibinfo {author} {\bibfnamefont
  {D.}~\bibnamefont {Elkouss}},\ and\ \bibinfo {author} {\bibfnamefont
  {S.}~\bibnamefont {Wehner}},\ }\bibfield  {title} {\bibinfo {title}
  {Netsquid, a network simulator for quantum information using discrete
  events},\ }\bibfield  {journal} {\bibinfo  {journal} {Communications
  Physics}\ }\textbf {\bibinfo {volume} {4}},\ \href
  {https://doi.org/10.1038/s42005-021-00647-8} {10.1038/s42005-021-00647-8}
  (\bibinfo {year} {2021})\BibitemShut {NoStop}%
\bibitem [{\citenamefont {Vardoyan}\ \emph {et~al.}(2021)\citenamefont
  {Vardoyan}, \citenamefont {Guha}, \citenamefont {Nain},\ and\ \citenamefont
  {Towsley}}]{Vardoyan_2021}%
  \BibitemOpen
  \bibfield  {author} {\bibinfo {author} {\bibfnamefont {G.}~\bibnamefont
  {Vardoyan}}, \bibinfo {author} {\bibfnamefont {S.}~\bibnamefont {Guha}},
  \bibinfo {author} {\bibfnamefont {P.}~\bibnamefont {Nain}},\ and\ \bibinfo
  {author} {\bibfnamefont {D.}~\bibnamefont {Towsley}},\ }\bibfield  {title}
  {\bibinfo {title} {On the stochastic analysis of a quantum entanglement
  distribution switch},\ }\href {https://doi.org/10.1109/tqe.2021.3058058}
  {\bibfield  {journal} {\bibinfo  {journal} {IEEE Transactions on Quantum
  Engineering}\ }\textbf {\bibinfo {volume} {2}},\ \bibinfo {pages} {1–16}
  (\bibinfo {year} {2021})}\BibitemShut {NoStop}%
\bibitem [{\citenamefont {Li}\ \emph {et~al.}(2023)\citenamefont {Li},
  \citenamefont {Fu}, \citenamefont {Liu}, \citenamefont {Xie}, \citenamefont
  {Li}, \citenamefont {Zhou}, \citenamefont {Yin},\ and\ \citenamefont
  {Chen}}]{Multiplexing1}%
  \BibitemOpen
  \bibfield  {author} {\bibinfo {author} {\bibfnamefont {C.-L.}\ \bibnamefont
  {Li}}, \bibinfo {author} {\bibfnamefont {Y.}~\bibnamefont {Fu}}, \bibinfo
  {author} {\bibfnamefont {W.-B.}\ \bibnamefont {Liu}}, \bibinfo {author}
  {\bibfnamefont {Y.-M.}\ \bibnamefont {Xie}}, \bibinfo {author} {\bibfnamefont
  {B.-H.}\ \bibnamefont {Li}}, \bibinfo {author} {\bibfnamefont {M.-G.}\
  \bibnamefont {Zhou}}, \bibinfo {author} {\bibfnamefont {H.-L.}\ \bibnamefont
  {Yin}},\ and\ \bibinfo {author} {\bibfnamefont {Z.-B.}\ \bibnamefont
  {Chen}},\ }\href@noop {} {\bibinfo {title} {Breaking universal limitations on
  quantum conference key agreement without quantum memory}} (\bibinfo {year}
  {2023}),\ \Eprint {https://arxiv.org/abs/2212.05226} {arXiv:2212.05226
  [quant-ph]} \BibitemShut {NoStop}%
\bibitem [{\citenamefont {Karp}(1972)}]{Karp}%
  \BibitemOpen
  \bibfield  {author} {\bibinfo {author} {\bibfnamefont {R.~M.}\ \bibnamefont
  {Karp}},\ }\bibinfo {title} {Reducibility among combinatorial problems},\ in\
  \href {https://doi.org/10.1007/978-1-4684-2001-2_9} {\emph {\bibinfo
  {booktitle} {Complexity of Computer Computations: Proceedings of a symposium
  on the Complexity of Computer Computations, held March 20--22, 1972, at the
  IBM Thomas J. Watson Research Center, Yorktown Heights, New York, and
  sponsored by the Office of Naval Research, Mathematics Program, IBM World
  Trade Corporation, and the IBM Research Mathematical Sciences Department}}},\
  \bibinfo {editor} {edited by\ \bibinfo {editor} {\bibfnamefont {R.~E.}\
  \bibnamefont {Miller}}, \bibinfo {editor} {\bibfnamefont {J.~W.}\
  \bibnamefont {Thatcher}},\ and\ \bibinfo {editor} {\bibfnamefont {J.~D.}\
  \bibnamefont {Bohlinger}}}\ (\bibinfo  {publisher} {Springer US},\ \bibinfo
  {address} {Boston, MA},\ \bibinfo {year} {1972})\ pp.\ \bibinfo {pages}
  {85--103}\BibitemShut {NoStop}%
\bibitem [{\citenamefont {Hartmanis}(1982)}]{ProofNP}%
  \BibitemOpen
  \bibfield  {author} {\bibinfo {author} {\bibfnamefont {J.}~\bibnamefont
  {Hartmanis}},\ }\bibfield  {title} {\bibinfo {title} {{Computers and
  Intractability: A Guide to the Theory of NP-Completeness (Michael R. Garey
  and David S. Johnson)}},\ }\href {https://doi.org/10.1137/1024022} {\bibfield
   {journal} {\bibinfo  {journal} {SIAM Review}\ }\textbf {\bibinfo {volume}
  {24}},\ \bibinfo {pages} {90} (\bibinfo {year} {1982})},\ \Eprint
  {https://arxiv.org/abs/https://doi.org/10.1137/1024022}
  {https://doi.org/10.1137/1024022} \BibitemShut {NoStop}%
\bibitem [{\citenamefont {Song}\ \emph {et~al.}(2012)\citenamefont {Song},
  \citenamefont {Wen}, \citenamefont {Guo},\ and\ \citenamefont
  {Tan}}]{finite}%
  \BibitemOpen
  \bibfield  {author} {\bibinfo {author} {\bibfnamefont {T.-T.}\ \bibnamefont
  {Song}}, \bibinfo {author} {\bibfnamefont {Q.-Y.}\ \bibnamefont {Wen}},
  \bibinfo {author} {\bibfnamefont {F.-Z.}\ \bibnamefont {Guo}},\ and\ \bibinfo
  {author} {\bibfnamefont {X.-Q.}\ \bibnamefont {Tan}},\ }\bibfield  {title}
  {\bibinfo {title} {Finite-key analysis for measurement-device-independent
  quantum key distribution},\ }\href
  {https://doi.org/10.1103/PhysRevA.86.022332} {\bibfield  {journal} {\bibinfo
  {journal} {Phys. Rev. A}\ }\textbf {\bibinfo {volume} {86}},\ \bibinfo
  {pages} {022332} (\bibinfo {year} {2012})}\BibitemShut {NoStop}%
\bibitem [{\citenamefont {Goldberg}\ \emph {et~al.}(1989)\citenamefont
  {Goldberg}, \citenamefont {Tardos},\ and\ \citenamefont
  {Tarjan}}]{NetworkFlow}%
  \BibitemOpen
  \bibfield  {author} {\bibinfo {author} {\bibfnamefont {A.~V.}\ \bibnamefont
  {Goldberg}}, \bibinfo {author} {\bibfnamefont {{\'E}.}~\bibnamefont
  {Tardos}},\ and\ \bibinfo {author} {\bibfnamefont {R.}~\bibnamefont
  {Tarjan}},\ }\href@noop {} {\emph {\bibinfo {title} {Network flow
  algorithm}}},\ \bibinfo {type} {Tech. Rep.}\ (\bibinfo  {institution}
  {Cornell University Operations Research and Industrial Engineering},\
  \bibinfo {year} {1989})\BibitemShut {NoStop}%
\end{thebibliography}%

\appendix

\section{Simulation of the multiplexing protocol}
\label{sec:A}

To calculate the necessary parameters for the router rate and the secret key rate, we simulate the multipartite multiplexing protocol. To do so, the memories of each party are stored as an array that can take either values of 0 (for empty memories) or an integer $i \in \mathbb{N}$, depending on the number of rounds the qubit is already stored. All simulations are done in Python. The steps of the simulation are the following:
\begin{enumerate}
    \item Starting from the empty memory configuration, a decision is made for each memory in turn as to whether a qubit arrives and is stored or not. This happens randomly with a probability of $\eta$ (given by the probability for a qubit to arrive at the router). If a qubit is stored, the corresponding array entry is set from 0 to 1. If the qubit is still in memory at the end of the round, the entry in the following round is increased by one. In further rounds, a qubit is only sent if the memory is still empty. 
    \item Based on the memory configuration, the maximum matching is performed to decide which memories are involved in which GHZ measurement. Depending on the strategy chosen, as well as the number of parties and the connection length, the matching is implemented differently (see App. \ref{sec:B}). 

    The memory configuration determines how many GHZ measurements can be performed in each round (see Eq. (\ref{eq:maxk})). This results in the value for $l$.
    Furthermore, the probability of the average number of storage rounds $\text{Prob}[\delta_i]$ can be determined at this point considering each $\delta_i$ given by each qubit of the i$^{th}$ party. 
    \item The memories whose qubits were used in a GHZ measurement are set back to 0.
    Depending on whether a cutoff is defined, all qubits whose number of storage rounds has exceeded the cutoff are removed from the memories as well. 
    \item Using this new memory configuration, the next round $s + 1$ is started and the memory storage rounds are increased by one (if not empty).    
\end{enumerate}
The protocol is performed a fixed number of rounds. Additionally, the protocol is repeated several times, to get the average values for $l$ and $\text{Prob}[\delta_i](s)$. 

\section{The matching algorithm}
\label{sec:B}
 
\begin{figure}
    \centering
    \includegraphics[scale=0.17]{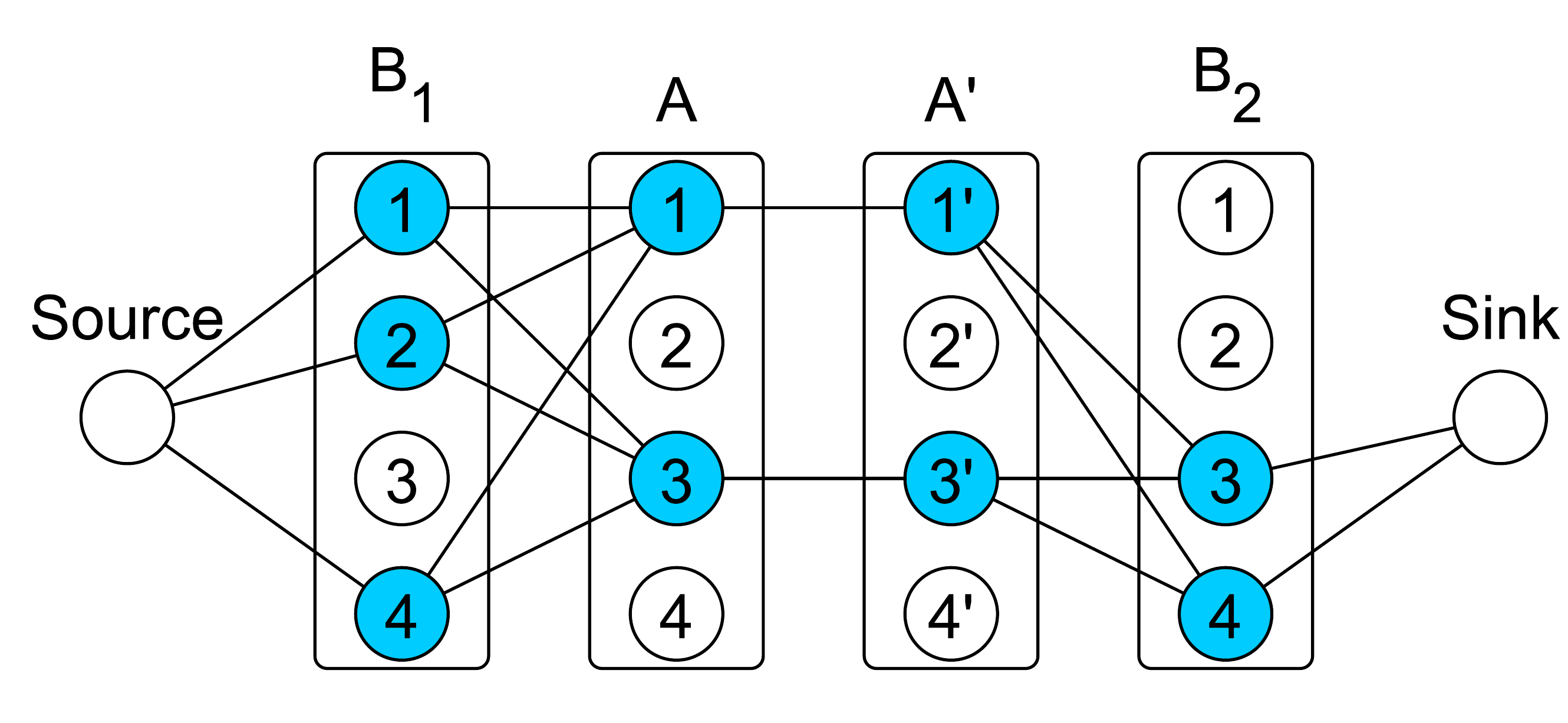}
    \caption{Representation of the matching problem for the tripartite graph as a network flow. A source and a sink are added to the graph. All edges have a capacity of one. To ensure that each party only appears once in a chosen flow (matching), it has to be ensured that each party only has one incoming or outgoing edge. Therefore, Alice's memory layer is copied and connected to the original layer accordingly. This ensures a single outgoing edge for A and a single incoming edge for A' resulting in the guaranty that A cannot be contained in two different flows (matchings). }
    \label{fig:Flow}
\end{figure}

The matching is realized in different ways depending on the strategy, the number of parties, and the connection length:

For the tripartite network, the matching problem can be reduced to the well-known Network Flow problem \cite{NetworkFlow}, as the graph structure given by the circuit in Fig. \ref{fig:GHZ} defines a flow from party B$_1$ over party A to party B$_2$. Assigning capacities to the edges and adding a source and a sink to the graph, the maximum flow from the source to the sink can be calculated. This corresponds to a maximum 3-dimensional matching. As each node is allowed to appear in only one matching, the capacity has to be set to 1. Additionally, the array of the party in the middle (here party A), has to be doubled (layer A'). Edges are only allowed to be drawn between a node and its own copy (see Fig. \ref{fig:Flow}). 
This has to be done, to assure, that memories are not chosen more than once, as the input flow has to equal the output flow for each node. For party B$_1$ and party B$_2$ this is guaranteed, since these nodes are only connected to the source or sink on one side, respectively. In Python, Network Flow can be realized using the \emph{maximum\_flow} function from \emph{scipy.sparse.csgraph}.

For networks with more than three parties, the matching is realized as follows. At first, all memories that are filled in one round are included in the set of nodes $V$ where the filled memories from one party form one subset $V_i$ of vertices. In the next step, all valid edges and the resulting hyperedges between the memories from the disjoint subsets are identified depending on the connection length $w$. Then, all possible combinations of hyperedges are considered sequentially. The first subset of hyperedges with no common vertices and maximum cardinality according to Eq. (\ref{eq:maxk}) that is found is chosen to be the matching. Note, that the complexity of the matching algorithm depends on the graph structure influenced by the maximal connection length. Only if $0<w<m-1$ (which corresponds to the finite-range multiplexing), this algorithm belongs to the class of $\mathcal{NP}$ problems. For parallel connections, i.e. $w=0$ there is only one matching that can be found, as no node has more than one incoming or outgoing edge, respectively. In the case of full-range multiplexing, i.e. $w=m-1$, a matching can be found by connecting the first filled memory from each party. Afterwards, these nodes have to be erased from the graph so that they cannot be chosen a second time. This is repeated until one party has no more filled memories. 

In the case of weighted matching, it is always necessary to find all allowed hyperedges (e.g. by \emph{combinations} from the Python package \emph{itertools}). From all combinations, the combination with maximum cardinality and maximum or minimum weight is chosen. This algorithm is also used for the tripartite network when weights are considered.

\section{Explicit calculations for the tripartite network}
\label{sec:Tripartite}

In the following, the calculations of the shared quantum state and the resulting quantum bit error rates are made explicitly for the case of the tripartite network. Starting from three Bell states $|\phi ^+\rangle = \frac{1}{\sqrt{2}} \left(|00 \rangle + |11 \rangle \right)$ held by A, B$_1$, and B$_2$, the input state for the GHZ measurement can be calculated. We assume that the qubits stored in the memory undergo depolarization, such that $|\phi^+ \rangle \langle\phi^+ |_i \rightarrow \rho_i^{dep} = F_i |\phi ^+\rangle \langle \phi^+| + \frac{1-F_i}{3} \left( |\phi^-\rangle \langle \phi^-| + |\psi^+\rangle \langle \psi^+| + |\psi^-\rangle \langle \psi^-| \right)$ for $i\in \{A, B_1, B_2\}$. The input state is then given by the product state 
\begin{align}
    \rho_{A_1B_1B_2}^{dep} &= \rho_{A_1}^{dep} \otimes \rho_{B_1}^{dep} \otimes \rho_{B_2}^{dep}  \nonumber \\
    &= \rho_{a^{(1)}a^{(2)}}^{dep} \otimes \rho_{b^{(1)}_1b^{(2)}_1}^{dep}  \otimes \rho_{b^{(1)}_2b^{(2)}_2}^{dep} 
\end{align}
The GHZ measurement is performed on the second qubit of the three parties ($a^{(2)}, b_1^{(2)}$, and $b_2^{(2)}$) according to the quantum circuit given in Fig. \ref{fig:GHZ}. 
The measurements performed at the end of the circuit are done in the $Z$-basis. In the case of measuring three times a zero, the remaining first qubits ($a^{(1)}, b_1^{(1)}$, and $b_2^{(1)}$), held by each party, are projected onto the $|GHZ_0^+\rangle$ state (with a certain fidelity $\Tilde{F}\leq1$ because of the depolarization). Depending on the measurement outcome, a projection onto another GHZ state is also possible. In this case, the desired $|GHZ_0^+\rangle$ state can be achieved by the parties changing their local qubit according to $Z^{m_a^{(2)}} \otimes X^{m_{b_1}^{(2)}} \otimes X^{m_{b_2}^{(2)}}$ with the announced measurement outcomes $m^{(2)}_{i}$ of the second qubit, where $i\in \{ a, b_1, b_2 \}$. The corrected final state resulting from a GHZ measurement of the depolarized initial states is then given by the following GHZ diagonal state:
\begin{widetext}
    \begin{align}
    \Tilde{\rho}_{A_1B_1B_2}^{dep} 
    = \lambda_0^+ |GHZ_0^+ \rangle \langle GHZ_0^+ | + \lambda_0^- |GHZ_0^- \rangle \langle GHZ_0^-| + \sum_{i=1}^3 \lambda_i \left( |GHZ_i^+ \rangle \langle GHZ_i^+|+|GHZ_i^- \rangle \langle GHZ_i^-|\right)
\end{align}
\end{widetext}
\clearpage
\newpage
with the GHZ states $| GHZ_i^{\pm} \rangle = \frac{1}{\sqrt{2}} \left( |i \rangle \pm |N^2 -1-i \rangle \right)$ and $i=0,1,\hdots, 2^{N-1}$ given in binary notation.\\
The GHZ diagonal elements $\lambda_i^{\pm}$ are given by: 
\begin{widetext}
\begin{align}
    \lambda_0^+ =& \hspace{0.1cm} \frac{1}{27} \left(4- F_A - F_{B_1} - F_{B_2} - 2F_AF_{B_1} - 2F_{B_1}F_{B_2} - 2F_AF_{B_2} + 32F_AF_{B_1}F_{B_2} \right) \\
    \lambda_0^- =& \hspace{0.1cm} \frac{1}{27} \left( 5 - 5F_A - 5F_{B_1} - 5F_{B_2} + 14F_AF_{B_1} + 14F_{B_1}F_{B_2} + 14F_AF_{B_2} - 32 F_AF_{B_1}F_{B_2} \right) \\
    \lambda_1 =& \hspace{0.1cm} \frac{1}{9} \left( F_{B_2} + 2F_AF_{B_1} -2F_AF_{B_2} - 2F_{B_1}F_{B_2} + 1\right) \\
    \lambda_2 =& \hspace{0.1cm} \frac{1}{9} \left( F_{B_1} - 2F_AF_{B_1} -2F_{B_1}F_{B_2} + 2F_AF_{B_2} + 1\right) \\
    \lambda_3 =& \hspace{0.1cm} \frac{1}{9} \left( F_A - 2F_AF_{B_1} - 2F_AF_{B_2} + 2F_{B_1}F_{B_2} + 1 \right)
\end{align}
\end{widetext}
with initial fidelity $F_i = \frac{1}{4}+\frac{3}{4}\cdot e^{-\delta_i/\tau}$ for each party $i \in \{ A, B_1, \hdots, B_{N-1} \}$. Note, that one of these fidelities has to be one in each round since we maximize the number of GHZ measurements $l$. 
The fidelity of the state after correction according to the measurement outcome is: 
\begin{align}
\label{eq:fidi}
    \Tilde{F}_{\Tilde{\rho}^{dep}} =& \hspace{0.1cm} \langle GHZ_0^+ | \Tilde{\rho}_{a^{(1)}b_1^{(1)}b_2^{(1)}}^{dep} | GHZ_0^+ \rangle \nonumber \\ 
    =& \hspace{0.1cm}  \lambda_0^+
\end{align}
For the QBER in X-basis, it holds
\begin{align}
\label{eq:Qex}
    Q_X &= \frac{1-\langle X ^{\otimes N}\rangle}{2} \nonumber \\
    &= \frac{1-\left(\lambda_0^+-\lambda_0^-\right)}{2} .
\end{align}
The bipartite QBERs are given by 
\begin{align}
\begin{split}
\label{eq:Qbi}
    Q_{AB_1} &= 2 \left(\lambda_2 + \lambda_3\right)   \\
    Q_{AB_2} &= 2 \left(\lambda_1 + \lambda_3\right) 
\end{split}
\end{align}
With the quantum bit error rates 
from Eq. (\ref{eq:Qex}) and Eq. (\ref{eq:Qbi}), the total error ($Q_X^{tot}, Q_{AB_i}^{tot}$) over the protocol (all rounds up to a current round $s_c$) between all outcomes of A and B$_i$ can then be calculated following Eq. (\ref{eq:totalQ}):
\begin{widetext}
    \begin{align}
        Q^{tot}(s_c) = \frac{\sum_{s=1}^{s_c} \langle l \rangle(s) \sum_{\delta_{a^{(2)}}}^{s} \sum_{\delta_{b_1^{(2)}}}^{s}\sum_{\delta_{b_2^{(2)}}}^{s} Q\left( \delta_{a^{(2)}}, \delta_{b_1^{(2)}}, \delta_{b_2^{(2)}} \right) \text{Prob}\left[ \delta_{a^{(2)}} \right] \text{Prob}\left[ \delta_{b_1^{(2)}} \right] \text{Prob}\left[ \delta_{b_2^{(2)}} \right](s)}{\sum_{s=1}^{s_c} \langle l \rangle(s)} 
    \end{align}
\end{widetext}
With this, we can finally compute the asymptotic secret fraction from Eq. (\ref{eq:rinfty}) and further the secret key rate given in Eq. (\ref{eq:KeyRate}).

\end{document}